\documentclass{iopart}

\usepackage{graphicx}

\usepackage[utf8]{inputenc}
\usepackage{textcomp}

\expandafter\let\csname equation*\endcsname\relax
\expandafter\let\csname endequation*\endcsname\relax

\usepackage{amsmath}

\usepackage[sort,compress,numbers,square]{natbib}

\usepackage{xcolor}

\newcommand*\reprint[2][Reprinted]{#1 with permission from~\cite{#2}.}
\newcommand*\reprintaps[2]{\cite{#1} Copyright (#2) by The American Physical Society.}
\newcommand*\reprintacs[3][Reprinted]{#1 with permission from~\cite{#2}.
Copyright (#3) American Chemical Society.}
\newcommand*\reprintelsevier[3][Reprinted]{#1 from~\cite{#2}.
Copyright (#3), with permission from Elsevier.}
\newcommand*\reprintaaas[2][Reprinted]{From~\cite{#2}.
#1 with permission from AAAS.}

\renewcommand{\exp}{\mathrm{e}}

\DeclareFontFamily{U}{euc}{}
\DeclareFontShape{U}{euc}{m}{n}{<-6>eurm5<6-8>eurm7<8->eurm10}{}%
\DeclareSymbolFont{AMSc}{U}{euc}{m}{n} 
\DeclareMathSymbol{\umu}{\mathord}{AMSc}{"16}

\renewcommand{\textmu}{$\umu$}

\newcommand*\GH[1]{#1}

\begin{document}

\topical{Nucleation and growth of thin films of rod--like conjugated molecules}

\author{Gregor Hlawacek}

\address{Physics of Interfaces and Nanomaterials, MESA+ Institute for
Nanotechnology, University of Twente, P.O. Box 217, 7500AE Enschede, The
Netherlands}
\ead{g.hlawacek@utwente.nl}

\author{Christian Teichert}
\address{Institute for Physics, Montanuniversitaet Leoben, Franz Josef
   Stra\ss{}e 18, 8700 Leoben, Austria}

\begin{abstract}
   Thin films formed from small molecules rapidly gain importance in
   different technological fields. To explain their growth,
   methods developed for zero--dimensional atoms as the film forming
   particles are applied. However, in organic thin film growth the
   dimensionality of the building blocks comes into play. Using the special case of the model
   molecule para--Sexiphenyl, we will emphasize the challenges that
   arise from the anisotropic and one--dimensional nature of building blocks. 
   Differences or common features with other rodlike molecules will be
   discussed. The typical morphologies encountered for this group of
   molecules and the relevant growth modes will be investigated. Special
   attention is given to the transition between flat lying  and upright
   orientation of the building blocks during nucleation.
   We will further discuss methods to control the molecular orientation and
   describe the involved diffusion processes qualitatively and
   quantitatively.

\end{abstract}

\submitto{\JPCM}

\maketitle

\section{Introduction}
\label{sec:intro}

Research over the last decades on growth processes at an atomic scale has
greatly enhanced our understanding of thin film formation and crystal
growth. In particular the realization that in addition to thermodynamic
effects also kinetic limitations at the surface play an important role during
the growth of thin films and crystals has helped to explain many 
growth
phenomena~\cite{Markov2003,Michely2004,Brune1998,Ratsch2003,Venables1984}. Although a great level of understanding has
been reached for many different processes, the vast majority of the systems
contained single atoms as the film forming entity. These are as
such zero--dimensional particles. 

\begin{figure}[tb]
   \begin{center}
      \includegraphics[width=8cm]{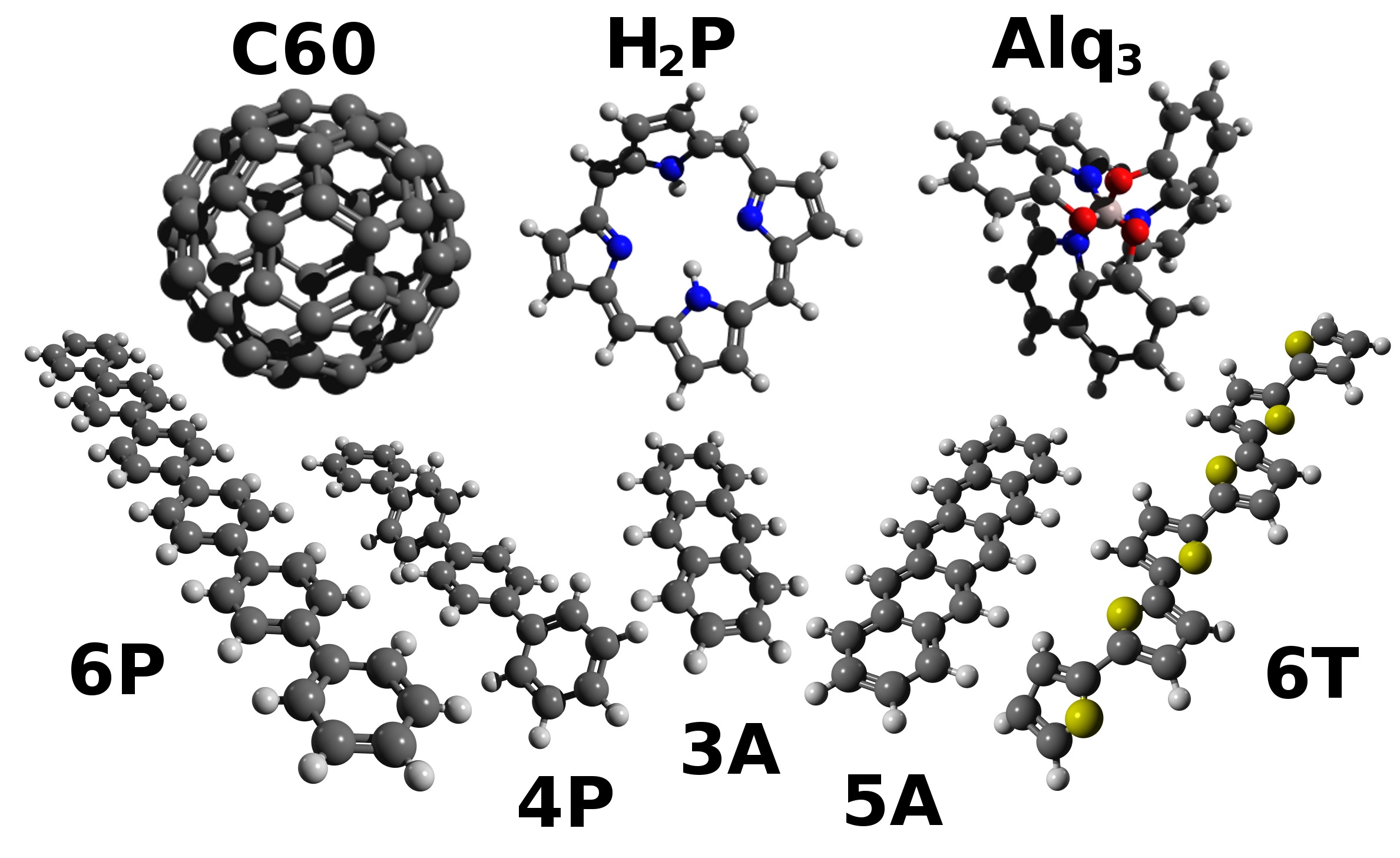}
   \end{center}
   \caption{Examples of organic semiconductor molecules. In the first row
      the oligomers C$_{60}$, Porphyrin (H$_2$P), and Alq$_3$
      (tris(8--hydroxyquinoline)aluminium) are shown. In the bottom row
      examples for some rodlike oligomers are presented. From left
   to right: para--Sexiphenyl (6P), para--Quaterphenyl (4P), Anthracene (3A),
   Pentacene (5A), and Sexitiophene (6T). Please note that while
   4P is shown in the twisted gas phase configuration, 6P is depicted flat as
   it is found in the bulk crystal structure.}
   \label{fig:org-mol}
\end{figure}

A new class of thin film materials---conjugated molecules
(see \fref{fig:org-mol} for examples)---has emerged in the
past 20 years. 
As is demonstrated in \fref{fig:org-mol}, such molecules can be
three--dimensional like tris(8--hydroxyquinoline)aluminium (Alq3), two--dimensional like the essentially planar
Porphyrins and Phthalocyanins or one--dimensional like the Acences,
oligo--Phenylenes and Tiophenes. The spherical C$_{60}$ molecule can either be
viewed as a large zero--dimensional or a
isotropic three--dimensional particle. Their use as organic semiconductors opens
exiting new possibilities for electronic and optoelectronic devices, in
particular flexible or stretchable~\cite{Wagner2012} ones. There is ample of evidence that
models used in inorganic epitaxy might also be applicable to this new class of
materials~\cite{Campione2007a,Ruiz2003a,Stadlober2006}. However, on a molecular level these
materials do
not always follow the well established findings of classical
epitaxy~\cite{Yang2008,Ruiz2004}. In this topical review, the
underlying reasons and consequences for thin film growth will be discussed.
A vast variety of different conjugated molecules
are currently under investigation and their number keeps steadily
increasing. As many of the effects described here are thought to be generic,
we will limit this topical review to rodlike molecules. They are a
representative group within the vast number of molecules with extended
dimensionality. In fact,
we will restrict ourself mostly to the oligophenylene molecule para--Sexiphenyl
(6P)~\cite{Fichou1999,Horowitz1991,Resel2003,bakandfra93}. This model
molecule is widely investigated for its potential use in organic thin film
transistors (OTFT), organic light emitting diodes (OLED) and solar cells. In
these devices, a specific molecular orientation is essential for optimal
performance.
By comparison to other rodlike molecules, we will highlight the
modifications that arise in the growth behaviour by changing molecular properties.

\begin{figure}[tb]
   \begin{center}
      \includegraphics[width=8cm]{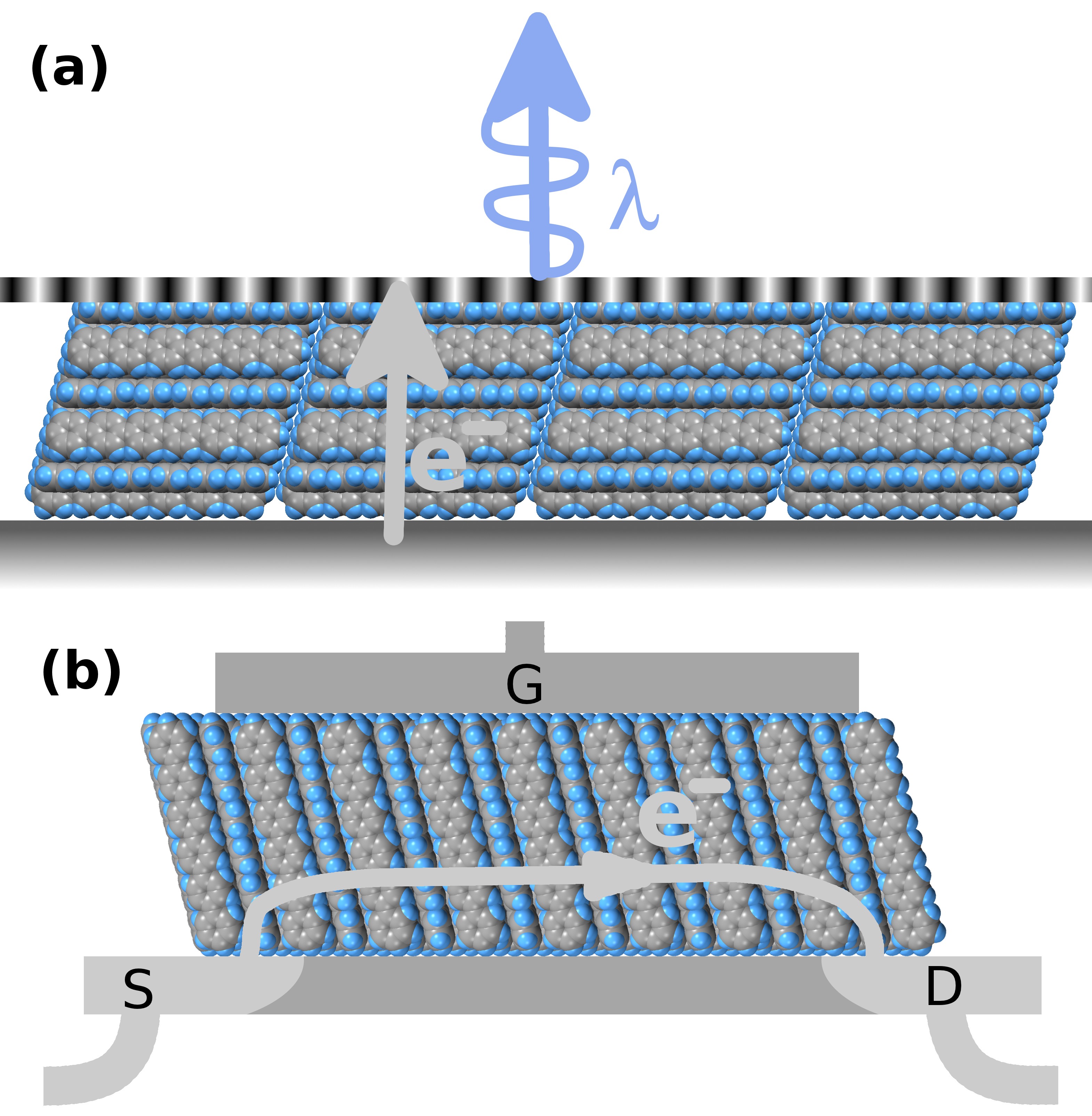}
   \end{center}
   \caption{Schematic comparison of required molecular orientations for device
   architectures based on rodlike conjugated molecules, using 6P as an
   example. (a) For OLED
   applications, the $\pi$--system should be
   oriented parallel to the substrate and the top electrode. For rod-- or
   plate--like molecules, this usually requires a flat lying configuration.
   (b) In an OTFT, the conjugated molecules should be oriented in an upright
   orientation. This facilitates an isotropic electric transport from source
   to drain parallel to the gate electrode. For holes the requirements on
   molecular orientation are
   unchanged as only the direction but not the path
   of the charge transport changes.}
   \label{fig:OLED-OTFT}
\end{figure}

After a brief introduction to small organic molecules, first the nucleation behaviour
will be discussed. It is during this initial growth stage when the
molecular orientation of the thin film is determined. This topic is of
special importance as different applications require different molecular
orientations on the substrate. The need to control this orientation is rooted in the
anisotropic properties of the molecules. In particular, the charge carrier
mobility for different crystallographic directions often shows a pronounced
anisotropy~\cite{DeBoer2004,Hu2008}.
As a rule of thumb, charge transport is
always best in directions with maximum $\pi$/$\pi$ overlap. Often this is
perpendicular to the long molecular axis or the molecular plane containing
the conjugated $\pi$--system for two- and three--dimensional molecules. In
\fref{fig:OLED-OTFT}, the required molecular orientations for a OTFT and an
OLED are depicted for the model molecule 6P. 
As can be seen, the desired orientation for an OLED application would be flat
lying molecules with their $\pi$--systems parallel to the electrodes. Charge
transport in such a device is then perpendicular to the $\pi$--system and the
long molecular axis or largest plane. Also the desired light emission is
maximised in such a configuration. On the other hand, for an OTFT application
an upright standing molecular configuration is wanted  to facilitate charge
transport parallel to the substrate in an isotropic way. Therefore,
possibilities to influence the nucleation behaviour and consequently the
resulting film morphology as well as the molecular orientation will be discussed. 

\begin{figure}[tb]
   \begin{center}
      \includegraphics[width=8cm]{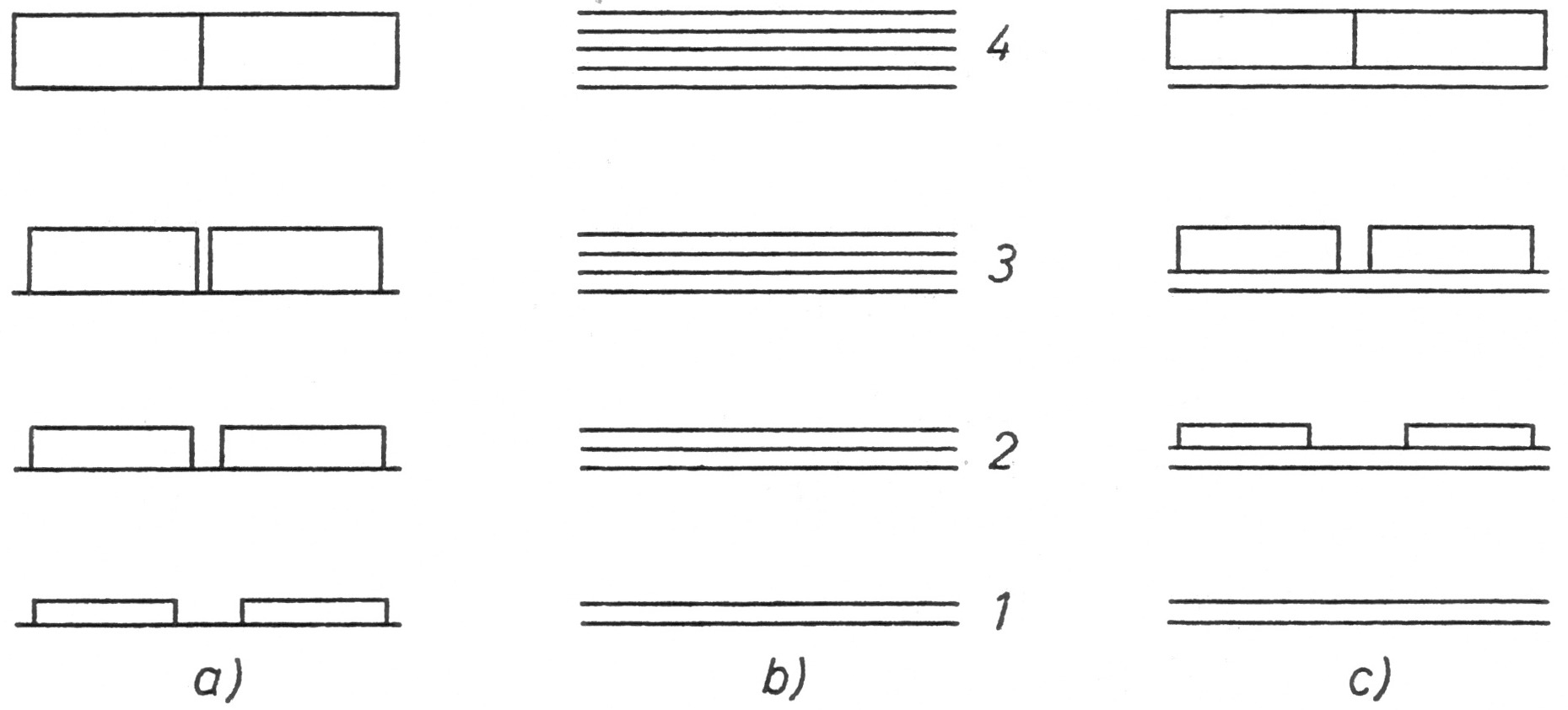}
   \end{center}
   \caption{Schematic growth morphology for (a) Vollmer--Weber, (b) Frank--van
      der Merwe, and (c) Stranski--Krastanov growth modes.
      \reprint{Bauer1958}}
   \label{fig:growth-modes}
\end{figure}

Secondly, we will describe different growth modes and mechanisms observed during the
formation of thicker films. 
Examples of Stranski--Krastanov and Vollmer--Weber growth
modes as well as Frank--van der Merwe or Layer--by--Layer (LbL)
growth will be presented~\cite{Bauer1958,Markov2003}.
Although difficult to achieve, the latter is often the desired growth mode to
fabricate continuous films with homogeneous thickness. The so
obtained smooth interfaces have a lower number of defects and yield generally a
higher charge carrier
mobility~\cite{Fritz2005,Steudel2004,Yan2010,Schumacher1982}. Which of the
above mentioned thermodynamical growth modes is realized, depends on the ratio of the
different surface free energies. The following requirement has to be
fulfilled for any thin film growth to happen.
The sum of the surface free
energy $\sigma_i$ between substrate and adsorbate, and $\sigma$ between the
adsorbate and the vapor are smaller then the surface free energy of the
substrate $\sigma_s$~\cite{Bauer1958}. Depending on the evolution of the
change of the surface free energy
\begin{equation}
   \Delta\sigma=\sigma+\sigma_i-\sigma_s
   \label{equ:surface-energy}
\end{equation}
during deposition, the three growth modes presented in \fref{fig:growth-modes} can be distinguished. For
$\Delta\sigma<0$ at all times it is feasible for the substrate to be covered
by the adsorbate layer. This growth mode is usually referred to as
Frank--van der Merwe, Layer--by--Layer (LbL) or two--dimensional growth. In case that
$\Delta\sigma>0$ at all times clustering will occur. This mode is called
Vollmer--Weber, island or three--dimensional growth. In case that
$\Delta\sigma<0$ for the initial deposited layers but changes to
$\Delta\sigma>0$ for subsequently deposited material, the film will start to
cluster after a thin uniform layer has been deposited. This growth mode is
called Stranski--Krastanov growth mode. The latter is frequently found in
inorganic heteroepitaxy for systems with a significant but not too large
mismatch~\cite{Teichert2002}.
Unfortunately in organic thin film
growth, the last two growth modes are much more common than the desired LbL
mode. 

The above sketched thermodynamic description of thin film growth is
not sufficient if the necessary diffusion processes are kinetically hindered.
A more atomistic approach that includes the actual
pathways is then needed to describe the observed morphologies
accurately~\cite{Teichert1994}. 
In particular, interlayer diffusion and the associated
barriers play a decisive role. Ehrlich--Schwoebel or step edge
barriers~\cite{Ehrlich1966,Schwoebel1966}, activation barriers for
intralayer diffusion,
and the anisotropy of these properties influence the final morphology.
Depending on the absolute and relative sizes of these barriers different
morphologies will be found in the resulting thick films. As a result of the
sizable step edge barrier often found in these molecular films, growth
phenomena like mound
formation and rapid roughening are frequently observed. As the strong van
der Waals interaction---typical for conjugated molecules---often dominates
all other intermolecular and molecule--substrate interactions,
three--dimensional growth is characteristic for organic semiconductor thin
films. 

For all
growth phenomena mentioned it is important to realize that already for
one--dimensional molecules at least two different scenarios have to be
distinguished. A smooth film grown in LbL mode might be useful for OLED
applications when formed from flat lying molecules. On the other hand, if the
molecules have an upright orientation, already a small number of layers
grown in LbL mode at the gate dielectric will yield a decent performance in
an
OTFT configuration. The reason that a small number of layers will suffice is
related to the fact that all important charge transport
processes are confined to the first two monolayers~\cite{Dodabalapur1995}. 

While many studies focus on the submonolayer
regime and interpret the behaviour of individual molecules, we follow a
mesoscopic approach. This is justified by the fact that the behaviour of
larger ensembles of molecules allows to infer information about the
molecular level
processes~\cite{Venables1984,Markov2003}. In addition, the investigated
mesoscopic size
range correlates well with the final device dimensions. The relevance of
different changes in the properties at this mesoscopic length scale can thus
directly be related to the device performance.

\subsection{Organic semiconductors}
\label{sec:intro:organic}

The conductivity of organic crystals has been studied already in the early
20$^{\mathrm{th}}$ century~\cite{Volmer1913,Koenigsberger1910}. However,
only with the discovery of electroluminescene these materials received additional
attention from the
semiconductor research community~\cite{Pope1963,Helfrich1965}. Inspired by
the Nobel prize awarded work of Alan Heeger, Hideki Shirakawa, and Alan
MacDiarmid in the 1970s, many researchers focused on conjugated polymers which
exhibit good conductivity if prepared properly~\cite{Chiang1977}. In the
1980s,
organic heterojunctions~\cite{Tang1986} and organic thin film
transistors~\cite{Horowitz1989,Koezuka1987,Burroughes1988} have been
demonstrated.  The final breakthrough happened after the realization of high efficiency
electroluminescence from organic light emitting diodes built both from
polymers~\cite{Burroughes1990,Braun1991} and
oligomers~\cite{Tang1987,Tang1989}. Nowadays, organic semiconductors are
either already used or are about to enter the market soon in countless
applications such as large--area--lightning, flexible solar cells, and
displays. These devices are based on the integrated use of OLEDs, OTFTs, sensors, and organic photovoltaic
cells~\cite{Reese2004,Samuel2004,Sariciftci2004,Borchardt2004,Brutting2005}.

With respect to their growth behaviour, organic semiconductors are of
interest for several reasons. As described above, classic surface science
treats zero--dimensional particles with a few exceptions such as Si
dimers~\cite{Swartzentruber1996}. The extended shape
of the used molecules does not only allow them to obtain different
orientations in space but also influences the way they interact with the
surrounding. It is the extended electronic system that is responsible for
the large intermolecular forces. The underlying van der Waals forces are
small for the individual constituting atoms but can add up to a few eV for the entire
molecule. The extended electronic system also helps to smoothen the
effective corrugation of the substrate. The molecule \textit{averages} over
many possible atomic adsorption sites to find the molecular adsorption
site with the minimum energy. Directly related to this is a large number of internal vibrational
degrees of freedom~\cite{Paserba2001,Fichthorn2002}. These have to be considered when discussing the
interaction at interfaces. These interfaces can occur between the
condensed phase and the 2D or 3D
gas phase, but can also be boundaries between different crystalline
domains.
Unfortunately, the large size of the building blocks results in a large number
of translational domains. Together with the usually low symmetry of the
crystal structure this can lead to additional disorder in the film, which
in turn hampers the final device performance.  For a general overview on the
properties of organic semiconductors,
the reader is referred to the following books~\cite{Brutting2005,Woll2009,Klauk2006}.

\subsection{Rod--like conjugated molecules}
\label{sec:intro:rod}

The three most important groups of rod--like conjugated molecules are the 
para--n--phenyls and the groups of acens and thiophenes (see lower part of
\fref{fig:org-mol}). The first two
will be of particular interest here, as they show opposed properties in some
important characteristics. The n--thiophene molecules are chiral and exhibit
interesting chiral phenomena in thin layers but not in the 
bulk~\cite{Kiel2007}.

In this article, we will focus primarily on
para--Sexiphenyl
(6P)~\cite{Yanagi1997,Yanagi1997b,Resel2003,Al-Shamery2008}. It is important
to realize at this point that the acenes and para--n--phenyls have a very
different
stiffness of their backbone. The single bonds in the n--phenyls allow a
certain flexibility of the backbone as compared to the acenes which possess a
much stiffer backbone formed by two bonds. In the case of the n--phenyls, the
phenyl rings can
twist with respect to each other (shown for 4P
in \fref{fig:org-mol}). This twist is
observed for single molecules either on the surface or in the gas phase.
However, in a bulk crystal the n-phenyls obtain a flat configuration (as
shown for 6P in \fref{fig:org-mol})~\cite{guhandgra99,bakandfra93}.
Typically, rod--like molecules like the acenes or phenylenes form a so called
herringbone bulk structure. This packing motif is characterized by an alternating
left and right tilt of the molecular plane around the long molecular axis. As
a result, the long hydrogen terminated side of one molecules faces the flat 
side of the neighboring molecules where the $\pi$--system is located. This
configuration balances the quadrupol moment of the molecules most effectively.
It should be noted that this packing motif is very different for what is found for two--dimensional
plate--like molecules that favour a planar bulk
stacking.

\section{Forming a nucleus}
\label{sec:nucleus}

In this section, the important quantity of interest is the so called
critical nucleus size $i^*$. It is defined as the biggest number of
particles forming a cluster that will become stable by adding one more
particle. While there are various way's to extract this number, the use of
rate theories~\cite{Venables1984} in combination with scanning probe
techniques has been proven to be extremely successful. The central result of
the underlying theory can be summarized by the relation
\begin{equation} 
   N\sim\left(\frac{F}{\nu}\right)^\chi 
   \label{equ:scaling}
\end{equation}
where $N$ is the island number density, $F$ denotes the flux of incoming
particles in numbers of particles deposited per unit time and surface area.  The
scaling exponent $\chi=i^*/(i^*+2)$ holds the dependence on the critical
nucleus size.
\begin{equation}
   \nu=\nu_0\exp^{-E_D/(k_BT)}
   \label{equ:hopping-island}
\end{equation}
is the particle jump rate on the surface.
Here, $E_D$ is the energy barrier the particle has to overcome,
$T$ the temperature and $k_B$ the Boltzmann factor. The
pre--exponential factor $\nu_0=2k_bT/h$ is often referred to as the attempt frequency.
For all practical purposes, this is on the order of $10^{13}$\,$s^{-1}$ in
inorganic systems. As we will see later, for organic systems $\nu_0$ can deviate substantially from
this value~\cite{Fichthorn2002}. 

A second approach is based on the scaling hypothesis stating that the island
statistics in the steady state regime will depend---besides $\nu$ and
$F$---only on the coverage
$\Theta$ via the mean island size~\cite{Bartelt1992}. Using this assumption, it is possible to derive
the critical nucleus size although this is not very easy to apply in practice. This result has been
extended by introducing an additional scaling for
the capture
numbers~\cite{Bartelt1996}. However, most often the empirical
scaling function of Amar and Family~\cite{Amar1995} is used. Recently a similar
approach has been introduced by Pimpinelli and Einstein based not on the
island size distribution but on the capture zone size
distribution~\cite{Pimpinelli2007,Pimpinelli2010,Mulheran1996}. The presented scaling methods have the
benefit that they require less data, and often a single experiment or even a
single image can be sufficient to extract the critical nucleus size. A
review and more in depth analysis of the strong points and weaknesses of the
individual methods can be found in~\cite{Markov2003,Michely2004}.

The above considerations are only valid for low coverages after the initial transient
nucleation regime but before the coalescence occurs. This intermediate
regime is called the steady state nucleation regime.

An additional obstacle---for all three methods---results from the
fact that the molecules are anisotropic. As a result they can obtain different
in plane but also out of plane orientations. As we will see later, the
single shot methods also carry the risk to overlook interesting kinetic
behaviour with respect to deposition rate or growth temperature. 

\subsection{Obtaining the critical island size}
\label{sec:nucleus:obtainig}

In the case of complete condensation, i.e., when re-evaporation can be excluded, the
island number density $N$ can be written as~\cite{Venables1984}
\begin{equation}
   N\propto\left(\frac{F}{\nu}\right)^\chi\exp^{E_N/k_BT_D}.
   \label{equ:nucleation-density}
\end{equation}
Here, $T_D$ refers to
the temperature of the sample during growth---often called deposition
temperature. The energy parameter
\begin{equation}
   E_N=\frac{i^*E_D+E_i}{(i^*+2)} 
   \label{equ:activation_barrier}
\end{equation}
can be split further into the activation
barrier for diffusion $E_D$ and the binding energy of the critical nucleus
$E_i$. Provided sufficient data is available, a plot of $\ln N$ vs $\ln F$ allows the extraction of the critical
nucleus from the slope 
\begin{equation}
   \alpha_F=\chi=\frac{i^*}{i^*+2}. 
   \label{equ:slope-rate}
\end{equation}
Plotting the island density in an Arrhenius fashion as $\ln N$ vs. $1/T_D$, one
can extract again from the slope
\begin{equation}
   {\alpha_T}_D=\frac{i^*E_D+E_i}{\left( i^*+2 \right)k_B } 
   \label{equ:slope-temp}
\end{equation}
information on the involved energies. An assessment of the quality of the
recorded data and the relevance of the extracted data can be obtained by
comparing the results from the two different methods~\cite{Venables1984,Potocar2011}
\begin{equation}
   y_{0F}+\alpha_F\ln F=y_{0T}+\frac{{\alpha_T}_D}{T_D}
   \label{equ:rate-quality}
\end{equation}
with $y_{0R}$ and $y_{0T}$ being the y--intercepts of the two above mentioned
plots.

The above described rate equation based approach has been used extensively in the past to obtain information on
the critical nucleus size. Typically sizes between 2 and 4 are found for
different rodlike molecules. For 5A on SiO$_2$, a value of 3--4 is
reported~\cite{Stadlober2006,Tejima2004} similar to 2--3 reported for 6P on
disordered mica(0001) substrates~\cite{Potocar2011}. However, due to the
large experimental data set necessary, in many studies also scaling laws
are used to determine the critical nucleus size. Most of the results are obtained
by applying island size scaling~\cite{Amar1995}. However, capture zone
scaling~\cite{Pimpinelli2010,Pimpinelli2007,Mulheran1996} seems to provide more
reliable results~\cite{Lorbek2011,Conrad2008,Miyamoto2009} for some cases.
In agreement with rate theory, scaling laws typically
yield values between 2 and 3 for 6P~\cite{Lorbek2011,Potocar2011} and
somewhat higher values between 3 and 6 for
5A~\cite{Ruiz2003a,Ruiz2004,Stadlober2006} on SiO$_2$ or cyclohexane terminated
Si\{001\}~\cite{Meyer2004}. However, in particular for the growth of 5A care
has to be taken with respect to the applicability of these single shot methods
because of fractal growth morphologies~\cite{Meyer2001}. 

\begin{figure*}[tb]
   \begin{center}
      \includegraphics[width=16cm]{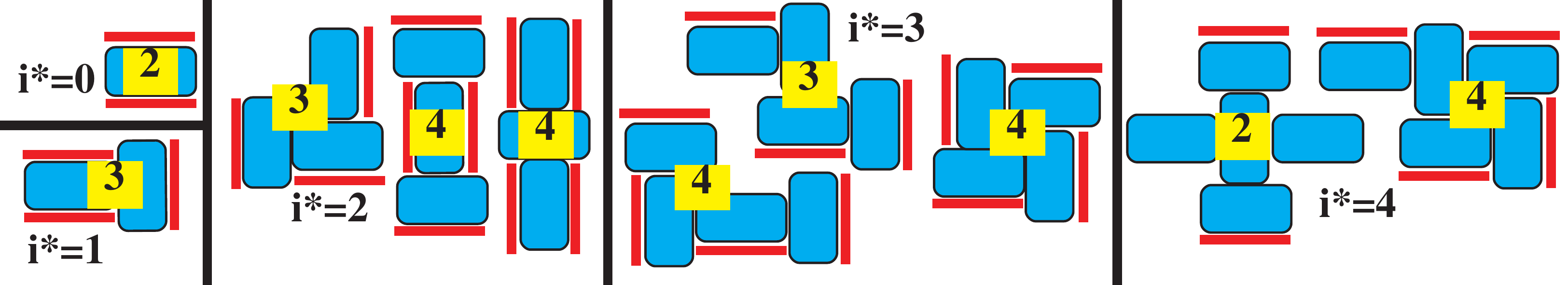}
   \end{center}
   \caption{Illustration of some of the possible configurations of the critical nuclei for a rodlike
   molecule in an upright configuration. Assuming a herringbone like configuration similar to the bulk
   structure, different possible nuclei configurations are sketched for $i^*$=0
   to $i^*$=4. The yellow labels give the number of fully exposed
   $\pi$--systems (red bars). Configurations with one and 5 molecules yield a minimal
   number of fully exposed $\pi$--systems. \reprint{Hlawacek2011a}}
   \label{fig:Crit-Nucl}
\end{figure*}

As will be demonstrated in the next 
section, the stability of the possible nuclei does
not necessarily depend in a homologous way on the nucleus size. In fact, due
to
effective shielding some configurations can be more stable than others (see
\fref{fig:Crit-Nucl}). 5A and 6P are very similar regarding
their herringbone packing in the bulk. Consequently, similar molecular
configurations in the critical nucleus will have comparable stability with
respect to other configurations.

As can be seen from  \eref{equ:hopping-island}, $\nu$
in \eref{equ:nucleation-density} depends also on the
attempt frequency $\nu_0$. From the y-intercept $y_{0T}$ in the above
mentioned Arrhenius plot, one can extract this quantity. It is important to
note that for molecules this value does not always correspond to the one
typically found in inorganic diffusion of zero--dimensional atoms
($\nu_0=1\times10^{13}$\,s$^{-1}$). In fact, here the value of $\nu_0$ can be much higher.
Values up $5.6\times10^{25}$\,s$^{-1}$~\cite{Mullegger2006} are reported
from thermal desorption spectroscopy (TDS). However, common values obtained
from TDS (experimental and theoretical) are in the order of
$1\times10^{17}$~\cite{Becker2006,Fichthorn2002,Fichthorn2007,Frank2010,Mullegger2006,Paserba2001,Tait2005,Winkler2009}.
Recent rate equation analysis yields a value of
$2\times10^{17}$ for the diffusion of 6P on mica~\cite{Potocar2011}. The explanation of these high values is
given by transition state theory~\cite{Zhdanov1991}. In this theory, the
pre--exponential factor depends on the partition functions of the particle in
the diffusive and adsorbed state. In contrast to an atom, a molecule
possesses
many vibrational and rotational degrees of freedom. In particular, the latter
ones contribute only to the partition function of the diffusive state. As a
result, the pre--exponential factor deviates from the well known
$1\times10^{13}$\,s$^{-1}$. The difference between values obtained by rate
theory~\cite{Potocar2011} and the sometimes extreme values obtained from
TDS~\cite{Mullegger2006} arises from the different target phases. 
While in a typical growth experiment---used for the rate equation approach
---the
molecule stays on the surface in a 2D gas phase, in a TDS experiment the
molecules enter the 3D vapour phase. However, the 3D phase has an even
higher number of degrees of freedom compared to the 2D gas phase, resulting in
different partition functions. 

\begin{figure}[tb]
   \begin{center}
      \includegraphics[width=8cm]{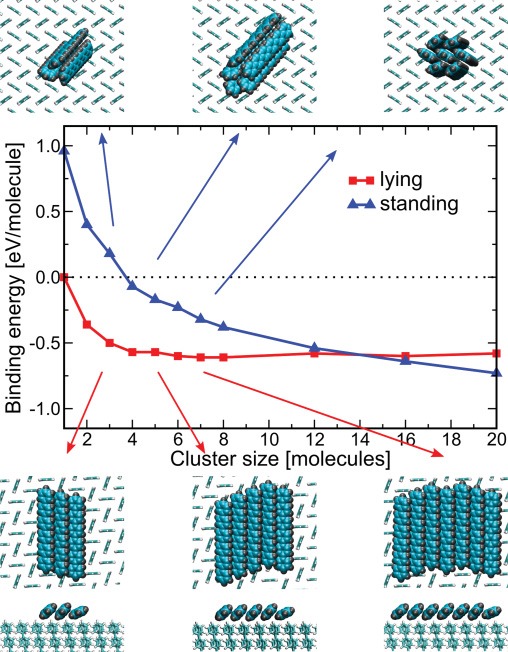}
   \end{center}
   \caption{6P cluster binding energy on a 6P(001) plane as function of
      cluster size. Graphs for clusters formed by upright standing and lying
      molecules are presented. Snapshots of possible cluster geometries are
      shown for cluster sizes of 3, 5, and 7 molecules. While the top row shows the
      clusters formed from upright standing molecules, the corresponding
      clusters formed by lying molecules can be found at the bottom. Please
      note that in the clusters formed by flat lying molecules all molecules
      have their $\pi$--system exposed. In the seven--molecule cluster---and to a
      certain extend in the cluster formed by 5 standing molecules---some
      molecules have their $\pi$--system saturated.~\reprintaps{Potocar2011}{2011}}
      \label{fig:clusters}
   \end{figure}

Independent of the above considerations, an interesting question concerning
islands formed by upright molecules remains unsolved at the moment.
How---and at what point during nucleation and growth---do the molecules
obtain an upright orientation? By considering the binding energy of an
upright
standing 6P molecule to the 6P(001) plane of 0.21,eV\footnote{private
communication P. Puschnig} and comparing it to the one for a flat lying
molecule of 1.27\,eV~\cite{Hlawacek2008} it is plausible to assume that a
single molecule will always obtain a flat lying
configuration~\cite{Goose2010}. Eventually flat lying molecules will meet
and form initially unstable dimers and trimers that will decay or continue to grow
and become stable when big enough. At some point---to form a film of
upright standing molecules---the molecules have to change from a lying to
upright standing configuration. Recent molecular dynamics (MD) simulations illustrate
the problem~\cite{Potocar2011}. The graph presented in \fref{fig:clusters}
shows the evolution of the cluster binding energy with increasing cluster
size. The cluster binding energy is calculated by comparing a situation with
$i$ molecules in a cluster to the same amount of $i$ molecules adsorbed in a
lying configuration on the 6P(001) surface. When the energy difference
between the cluster and the separate molecules is negative, the cluster is
thermodynamically 
stable. One can see that for $i<4$ only flat lying clusters are stable,
while clusters formed from upright standing molecules are entirely unstable.
For cluster size between 4 and 14 molecules, clusters of upright standing
molecules are stable but lying clusters would still be favored. Only for
clusters bigger than 14 molecules 6P continues to grow in the required (001)
orientation. However, the exact numbers will also depend on the details of
the interaction with the substrate. In the MD simulation presented in
\fref{fig:clusters}, the clusters rested on the 6P(001) surface rather than
on an amorphized mica substrate---discussed above---which is hard to
simulate. In any case, the simulation results will be correct for second
layer nucleation as well as
all following layers.

More information can be extracted from the graphs shown in
\fref{fig:clusters} on the growth of standing molecules. Already for
clusters of only two molecules ($i^*=1$) the energy gain is quite large
(on the order of $10\times kT$) and these clusters would be very stable. 
Furthermore, a simple geometric argument based on counting the number of fully exposed
$\pi$--systems arrives at the same critical island size of
$i^*\approx3\ldots5$ for upright standing
molecules. This is also the value often reported in
literature~\cite{Potocar2011,Tejima2004,Lorbek2011}. In \fref{fig:Crit-Nucl}, several possible configurations of
critical nuclei from $i^*$=0 to $i^*$=4 are sketched. The fully exposed
$\pi$--systems are also marked and their number is given. A large number of
exposed $\pi$--systems is
energetically unfavourable, and the system would in general try to minimize
their number. Besides the rough
nature of the model it turns out that only for clusters of four ($i^*=3$) or more
molecules the number of fully exposed $\pi$--systems becomes smaller than the
number of molecules. From this simple model we can conclude that once
the balance between molecules and exposed $\pi$--systems swings towards the
molecules the nuclei become stable. This is in good agreement with
the above presented MD results. The consequence of this can be seen twice in
\fref{fig:clusters}. First, the clusters formed by upright standing molecules
become stable around a size which allows for molecules that do not belong to
the island rim. Second, the
final slope of the two graphs is different. With increasing cluster size the
number of energetically more
favourable molecules which are not part of the rim increases faster for
clusters composed from upright standing molecules than for
the flat lying nuclei. At least for the investigated size range, clusters of
flat lying molecules contain only molecules which expose at least one
$\pi$--system to the vacuum. \GH{Nucleation processes
   involving small metastable clusters (e.g. non--epitaxial dimers) that eventually convert into larger
   stable clusters are also observed in
   inorganic semiconductor epitaxy~\cite{Schroeder1995,Filimonov2007,Filimonov2012}.}

The important question is, how do these flat lying molecules obtain an
upright standing orientation. Besides the reorientation of the whole cluster
other scenarios are in principle possible too. As we will see below, defects
on the substrate surface 
influence the orientation of the molecules. First principle calculations reveal
that already a single defect in an otherwise perfect surface can alter the
molecular orientation from flat lying to upright
standing~\cite{Tsetseris2005}. The initial flat
lying nucleus can act in a similar way. Molecules arriving later will undergo a
kind of defect nucleation and obtain an upright standing configuration. 
This is also observed in experiments.
When the coexistence of structures formed from flat lying and upright
standing molecules is observed, the first ones are often responsible for
nucleating the latter~\cite{Andreev2006,Fichou2007}. 

Furthermore, already small clusters formed from upright standing nuclei
should grow faster than their counterpart formed from flat lying molecules.
While molecules with all rotational orientations can be incorporated into a
cluster of standing molecules, they will eventually have to rotate in the
case of a cluster of flat lying molecules. In any case, the calculation only
yields the energetically most favourable configuration. It does not contain
information on the probability that it can actually form. Thus, although
energetically favourable the lying nuclei is actually more difficult to form
due to the rotational hindrance. \GH{However, this implies that no
   substantial energy barrier exists for the
reorientation of molecules from flat lying to upright standing. Although no
information is available on such a barrier, it can not be too
large, since the final morphology is dominated by upright standing molecules.
Such a morphology would be unlikely in a scenario where initially both orientations
compete and a high barrier would exist to obtain the upright orientation. In
such a competing scenario---where both types
of nuclei can form---a cluster of upright molecules could be kinetically
stabilized just because it can grow in size much easier.}
 Such an
attachment limited aggregation (ALA) for 6P has been observed and a critical nucleus
size of $i^*=7\pm2$ is found~\cite{Tumbek2012}. This is in reasonable
agreement with the numbers obtained in the above simulations for the
transition from unstable to stable for clusters formed by upright standing
molecules. In such a scenario, no reconfiguration of the cluster from lying
to upright would be necessary.

\subsection{Tuning molecular orientation and the role of defects}
\label{sec:nucleus:defining}

Despite the persisting problems with respect to nucleation, several
groups have succeeded in controlling the nucleation behaviour. This is an
important step towards realization of functional devices, since different
functionality requires different molecular orientations (see
section \ref{sec:intro} and \fref{fig:OLED-OTFT}). 

A convenient, but technically not very practicable, method to control
molecular orientation is via the substrate. The observed changes in orientation
go hand in hand with a change of the surface free energy of the
substrate. Although the surface free energy is therefore the most obvious
ordering parameter for the change from flat lying to upright standing, we
will use a different approach here. It is easy to see that the crystalline
structure of the substrate surface plays a crucial role for the
in--plane molecular orientation. However, the degree of order present in the
substrate surface can also be decisive with respect to an upright or flat
lying molecular orientation. In fact, the effect of changing the surface
structure can often dominate over the behaviour expected from a substrate
according to its other physical properties. A model system demonstrating
this is the deposition of para--Sexiphenyl onto muscovite mica surfaces.
Several groups
have shown that for a large number of rodlike molecules---such as 6P, 5A,
and 6T---nanofibers formed by flat lying molecules are the dominating
morphology on clean mica
surfaces~\cite{Akai-Kasaya2010,AAndreev2000,Teichert2006,Kankate2009,Balzer2005a,Al-Shamery2008,Al-Shamery2009}. 

\begin{figure}[tb]
   \begin{center}
      \includegraphics{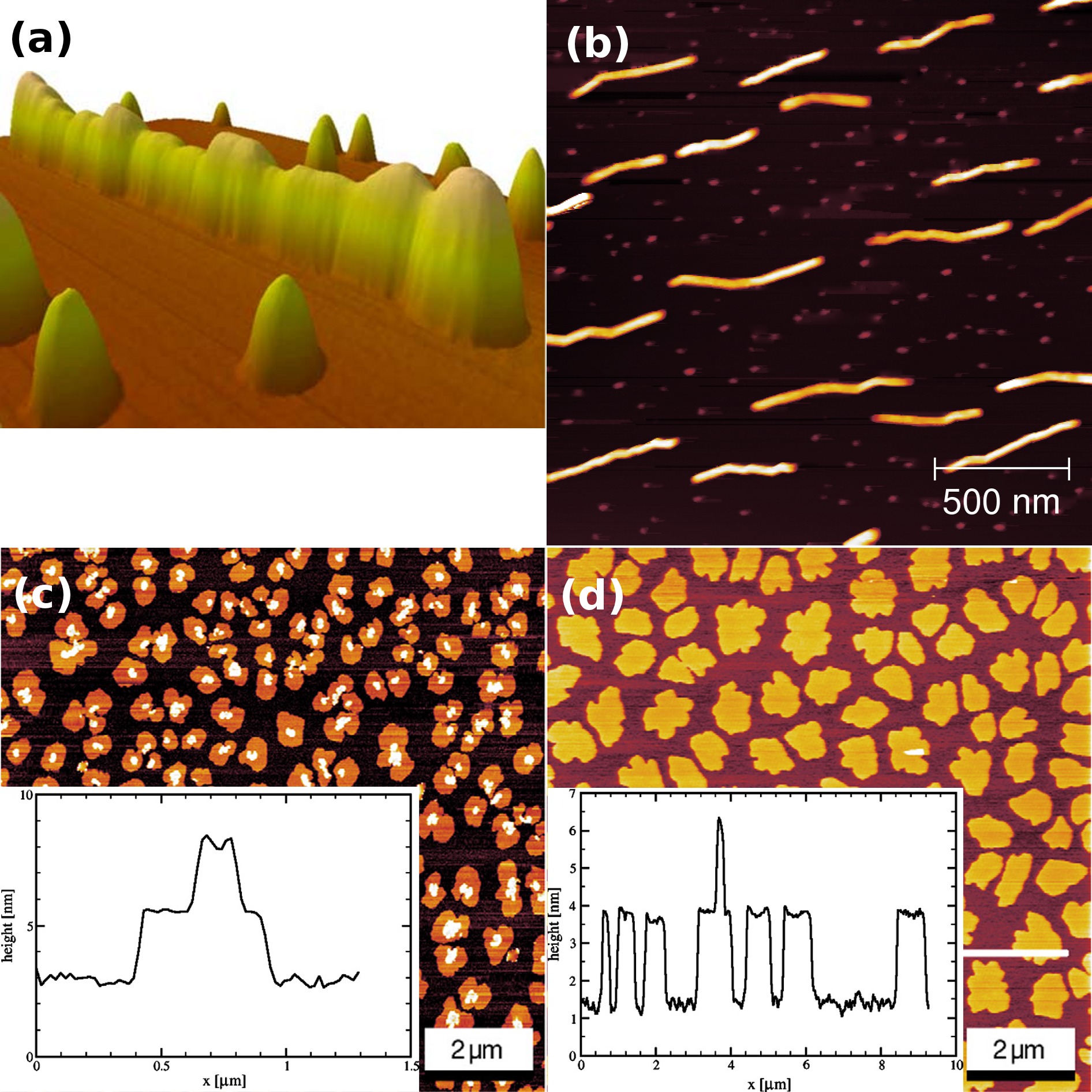}
   \end{center}
   \caption{AFM analysis of 6P on mica: (a) 3D representation of a 850\,nm
      long chain of 6P crystallites on mica. The film has been grown in
      35\,s at a
   substrate temperature of 360\,K. (b) OMBE grown 6P
   fibers on mica(0001). The film has a nominal thickness of 4\,nm, and the
   sample temperature during growth was held at 360\,K. The fibers consist of
   long segments. (c) 6P islands on carbon covered mica  grown at
   330\,K (nominal film thickness 1\,nm). (d) 6P islands
   on sputtered mica grown at 330\,K (nominal film thickness 1\,nm).
    The insets in (c,d) are cross sections revealing
   a terrace height of 2.6\,nm corresponding to the length of the molecule.
   (a): Reprinted from~\cite{Teichert2006}. With kind permission from
   Springer Science and Business Media. (c),(d):
   \reprintelsevier{Frank2007}{2007}}
   \label{fig:6PonMica}
\end{figure}

In \fref{fig:6PonMica}, typical morphologies obtained by Atomic Force
Microscopy (AFM) of 6P films grown on mica are presented.
\Fref{fig:6PonMica}(a) represents a particular interesting case of nanofiber
formation on mica(0001). Here, the spontaneous rearrangement of small
crystallites on top of a wetting layer into chains of crystallites is
observed~\cite{Teichert2006}. We will discuss this interesting growth later
in more detail. Changing
the growth method from the used High Vacuum (HV) Hot Wall Epitaxy (HWE) to Organic
Molecular Beam Epitaxy (OMBE) under Ultra High Vacuum (UHV) conditions
does not not influence the existence of a wetting layer. However, as can be seen from
\fref{fig:6PonMica}(b) this change in growth conditions results in the formation of larger more uniform
chain segments with a less pronounced internal
structure~\cite{Hlawacek2007,Frank2007,Balzer2002}. The above mentioned
rearrangement process happens in--situ during HWE growth. 

\begin{figure}[tb]
   \begin{center}
      \includegraphics[width=8cm]{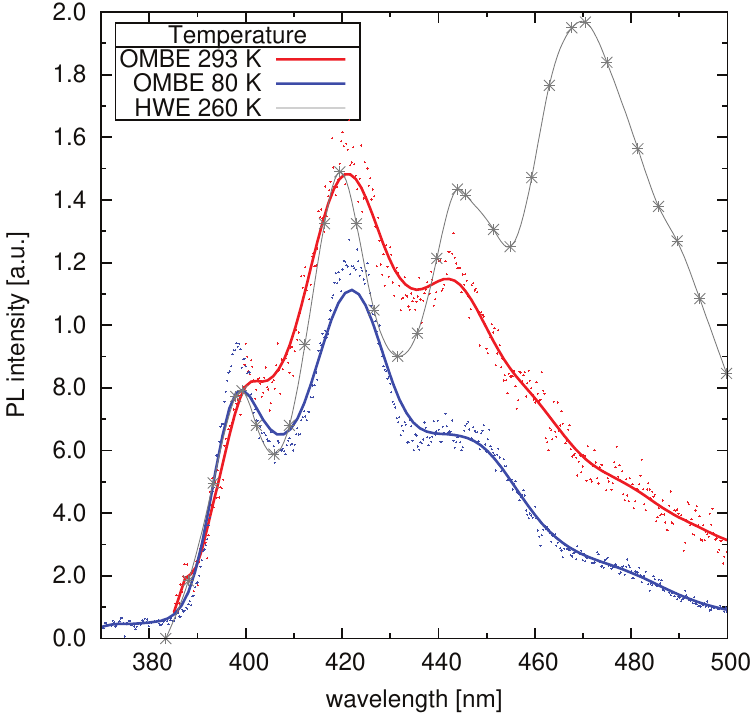}
   \end{center}
   \caption{(a) Steady state photoluminescence of UHV OMBE grown 6P fibers
   on mica (0001). The broad band emission around 480\,nm found in HWE grown
   fibers is missing. All curves are
   normalized to the 0-0 transition at 399\,nm. PL 
   measurements of OMBE grown films courtesy A. Kadashchuk (\reprint[Data reproduced]{Kadashchuk2004}). }
   \label{fig:mica-PL}
\end{figure}

The influence of the vacuum conditions can also be seen in photoluminescence
(PL) spectra obtained from these anisotropic fibers (\fref{fig:mica-PL}). 
In contrast to steady state PL spectra obtained from HV HWE grown 6P
fibers~\cite{Kadashchuk2004}, the broad band emission around 480\,nm is
suppressed for UHV OMBE grown fibers. The missing band at 480\,nm can be
connected to defect states resulting from structural defects. The absence of
this band suggests a higher quality of the OMBE grown films. 

The important conclusion of the above observation is the following. HV
conditions are usually sufficient to obtain reproducible and well performing
organic thin films. Given that the molecules are reasonable stable against
oxidation and UV light the obtained structures do not change over time in
ambient conditions. However, the morphology (and the resulting properties)
of organic thin films can easily be influenced by other small molecules.
Results obtained under (U)HV conditions should be carefully reviewed with
respect to 
their validity under ambient conditions.
In particular wetting layers and other possibly metastable structures far
away from the bulk structure are sensitive to adsorbates. 

\Fref{fig:6PonMica}(c) and (d) show the result of two different approaches
to reorient the molecules from flat lying to upright standing. The
molecules forming the crystallites and fibers in \fref{fig:6PonMica}(a,b)
have their long molecular axis parallel to the substrate. To obtain the
morphology presented in \fref{fig:6PonMica}(c) a
surfactant\footnote{The term surfactant is used here for a substance that
influences the growth.}---namely carbon---has been predeposited on the clean mica(0001) surface. TDS has revealed
that the full saturation coverage of surfactant
completely suppresses the formation of a wetting layer~\cite{Frank2007}. In the presence of
the surfactant, 6P films of upright standing molecules are formed. Similar
experiments have been performed with 6P~\cite{Mullegger2006}
and 
4P~\cite{Hlawacek2004,Resel2005,Mullegger2005} films on gold surfaces. In
both cases, a reorientation of the molecules has
been observed. It has to be pointed out that prior to the flat--lying/upright
transition the epitaxy between the molecular film and the substrate
is weakened. 
For the growth of 4P on Au\{111\} this is very well documented. On a
clean Au\{111\} surface, the (211) contact plane of 4P has all molecules
with their long axis parallel to the surface. Half of the molecules
have their $\pi$--system parallel to the surface. After the addition of only
15\% of a monolayer of carbon, the 4P contact plane changes to (201). The
long axis of the molecules is still parallel to the substrate surface.
However, in this 4P plane all molecules are tilted into an edge--on
configuration, where the $\pi$--system is not fully facing the substrate
anymore. At the same time, the in--plane molecules loose their 
alignment with the substrate. After a further increase of the surfactant
coverage to 0.5\,ML, the
molecules obtain an upright standing orientation. The new contact plane is
the 4P(001).
The weakened interaction with the substrate results in the
formation of bent 4P nano--fibers with a (201) contact plane on gold and
eventually the formation of mounds composed of upright standing
molecules~\cite{Hlawacek2004,Resel2005}.

Finally, \fref{fig:6PonMica}(d) also shows 6P islands on mica formed by
upright standing molecules. However, this morphology has been achieved by
breaking the surface symmetry of the substrate by ion bombardment of the
substrate~\cite{Frank2007}. The nominal
1\,nm thick film presented in \fref{fig:6PonMica}(d) consists of islands
formed by upright standing molecules as is demonstrated by the 
cross section presented in the inset. The same reorientation can also be
achieved by heating mica substrates to elevated
temperatures~\cite{Hlawacek2007}. More insight into the possible root cause
for the reorientation on mica has recently been obtained using TDS. Putsche
et al. demonstrated recently that the existence of a wetting layer and the final
orientation of the molecules is determined by the amount of potassium
present on the surface~\cite{Putsche2012}. The mechanism for the
reorientation is similar to what has been shown above for a carbon
predeposition.  

Although similar experiments exist for other small conjugated molecules,
the above series of experiments is unique as it clearly shows the importance
for a well ordered periodic substrate which can guide the molecules.
The gradual loss of this guidance initially leads to an inferior quality of
the crystals formed by flat lying molecules. Finally, after significantly
disturbing the 
surface order by surfactants or sputtering,
 the molecules choose to form films 
where their long molecular axis is perpendicular to the substrate surface. The loss
of order in the substrate can be the result of a surfactant layer,
sputtering, or the molecules themselves acting as
\textit{homosurfactants}. The latter is in particular true if
the molecules are adsorbed on the substrate in a hit--and--stick mode. The
initial molecules can not align in an ordered manner and create a disordered
substrate for the next layer~\cite{Cranney2009,Khokhar2010}. 
As expected from this discussion, using an amorphous substrate
such as SiO$_2$ will lead to the formation of films formed by upright
standing molecules~\cite{Conrad2009,Lorbek2011,Loi2005}. However, often the
layer of upright standing molecules itself can again act as a well ordered
substrate and can promote the successive growth of structures formed by lying
needles. See for example the needles forming on the 6P
mounds grown on sputtered mica, as presented in
\fref{fig:mound-morphology}(b,c)~\cite{Hlawacek2008}.

\begin{figure}[tb]
   \begin{center}
      \includegraphics[width=8cm]{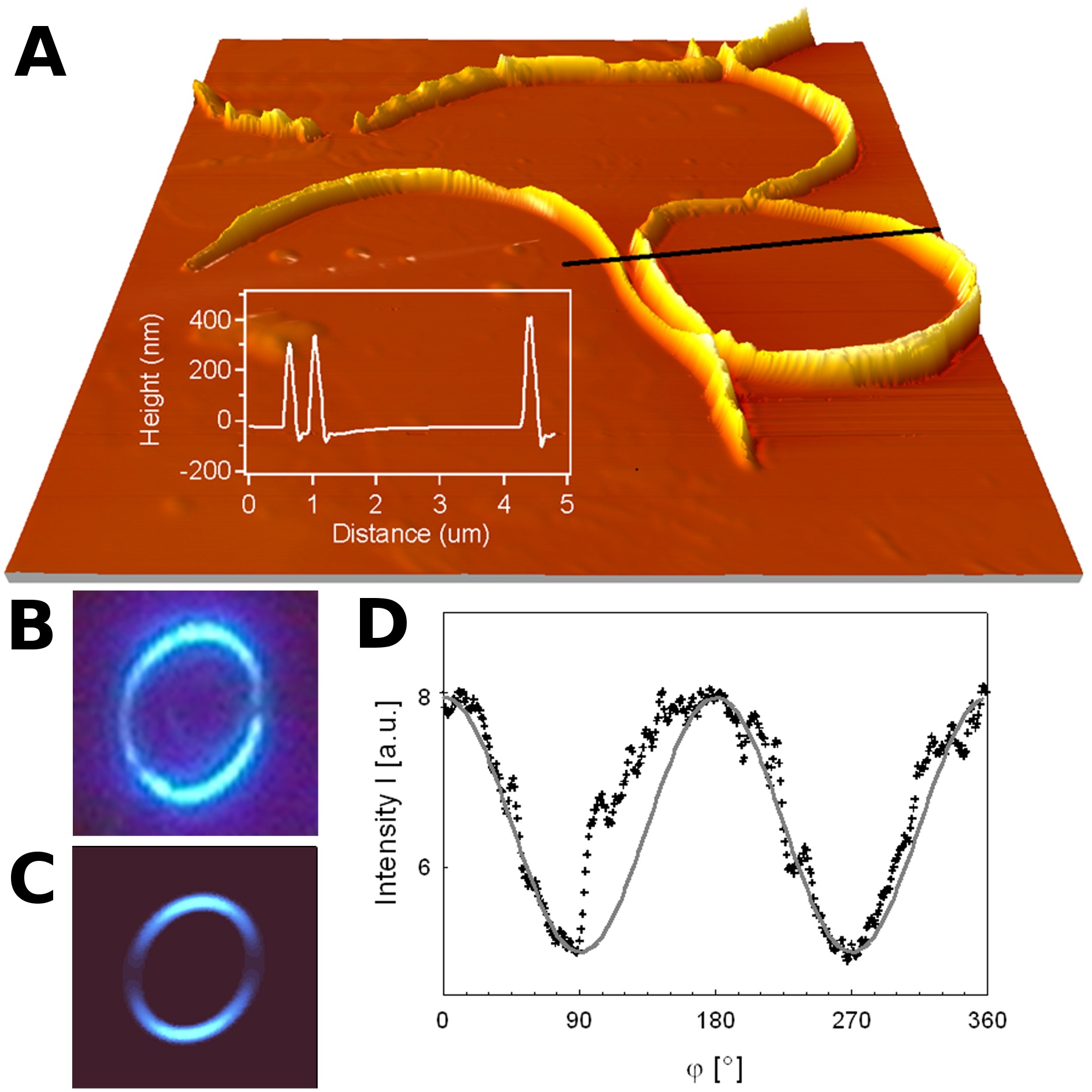}
   \end{center}
   \caption{(A) AFM image (10\,\textmu{}m$^2$ by 10\,\textmu{}m$^2$) showing a 6P
      nano ring on pretreated mica. The inset presents the cross section
      along the indicated line. (B) Luminescence micrograph and (C)
      corresponding simulated image. In (D), the intensity along the rim
      (symbols) is plotted together with the theoretical predicted one
      according to Malus' law. \reprintacs[Adapted]{Balzer2003a}{2003}}
   \label{fig:nano-rings}
\end{figure}

An extreme example of surfactant mediated growth of nanofibers~\cite{Balzer2003a} is presented in
\fref{fig:nano-rings}. 
The mica surface has been pretreated with either water or methanol to remove
the potassium atoms from the cleavage
surface~\cite{Nishimura1995,Nishimura1994}. It has been shown by AFM that
the so created rings are several hundred nanometer high and have a typical diameter of
4~\textmu{}m (see \fref{fig:nano-rings}(a)). The luminescence micrograph presented
in \fref{fig:nano-rings}(b) and the corresponding simulation
(\fref{fig:nano-rings}(c)) reveal two things. First, the molecules are lying
flat on the surface, and second their in--plane orientation rotates along the
ring. As only molecules having their long axis parallel to the polarization
of the incoming light yield maximum fluorescence, one observes a sinusoidal
change in intensity along the ring
(\fref{fig:nano-rings}(d))~\cite{Balzer2003a}.

\section{Diffusion and thin film growth}
\label{sec:diffusion}

In the previous section we have discussed the initial stages of film growth
i.e. the nucleation.
After islands with a certain orientation have formed they need to
grow, coalesce, and evolve into a complete film. Unfortunately, the
asymmetric building blocks used, often favor the formation of rough and
anisotropic surface
structures~\cite{Hlawacek2008,Kowarik2006,Durr2003,Mikami1998}. Typically,
one aims at a smooth film with a low number of grain boundaries and
a flat surface. Grain
boundaries represent inhomogeneities in the thin film and will have a
negative effect on the transport
properties~\cite{Verlaak2007,Verlaak2003,Horowitz1999a,Tkaczyk2004}. On top
of that they also affect the optical properties of the thin film
devices such as nano fiber wave
guides~\cite{Balzer2003,Balzer2003a,Andreev2006b,Cordella2007}. Different
growth techniques have been employed to achieve a smaller number of grain
boundaries. Notably the use of thin organic layers~\cite{Fritz2005} or Self
Assembled Monolayers (SAM)~\cite{Hill2009} and Super Sonic Molecular Beam
Deposition (SuMBD)~\cite{Wu2007} allow to influence the grain size without changing
the substrate.

As has been discussed in the introduction, there can be severe deviations
from the thermodynamic growth modes (\fref{fig:growth-modes}), if the necessary diffusion mechanisms
are kinetically hindered. The analysis of the morphology and if necessary structural information
allows to extract important physical quantities such as step edge barriers
as well as parameters describing the growth mechanism. The generalized
rules that can be deducted from the in--situ and ex--situ observations of film
growth are valid for many molecules.

\subsection{Mound formation}

As we have discussed earlier, thin films of conjugated organic molecules
which have their long axis perpendicular to the substrate are desired for
OTFT like applications. Although several approaches exist to achieve this
desired orientation, the resulting films are often characterized by the
formation of growth mounds due to rapid
roughening~\cite{Ruiz2004,Stadlober2006,Durr2003,Fritz2005,Zhang2009,Zhu2011}. The
resulting rough interface is undesired as it negatively affects the charge
carrier mobility~\cite{Schumacher1982,Fritz2005,Steudel2004,Yan2010}. 
Nevertheless, these rough films have been studied in the past as they
allow interesting insights into molecular diffusion processes and thin film
growth mechanisms. 

The process of rapid roughening is a result of an imbalance in the surface
diffusion over step edges. If the downward particle flux over the step edge
is dominating, the desired Layer--by--Layer growth mode is facilitated.
However, in the case of a high step edge barrier for downward
diffusion, which is described by a significant additional barrier---the
so called Ehrlich Schwoebel barrier
(ESB)~\cite{Schwoebel1966,Ehrlich1966}---mound formation will occur and
rough morphologies are the result. For the case of a very high
ESB and realistic growth conditions, only atoms landing on the terrace $h-1$ will be incorporated into
terrace $h$. The coverage $\Theta$ of terrace $h$ at time $t$ for a given
amount of deposited material $\overline{h}=Ft$ with the flux $F$ can than be
expressed by a Poisson distribution
\begin{equation}
   \Theta_h(t)=1-\exp^{-\overline{h}}\sum^{h-1}_{n=0}\frac{\overline{h}^n}{n!}.
   \label{equ:poisson-coverage}
\end{equation}
Because $\Theta_1=1-e^{-\overline{h}}$ will always be smaller than 1, the
first layer---and also all subsequent ones---never closes. As a result steep
trenches are observed. Such Poisson shaped mounds have been described in
inorganic systems
already early indirectly~\cite{Seah1972} and using real space
methods~\cite{Meinel1988}. An in--depth theoretical analysis has been
conducted later by Elkinani and Villain who used the ancient Greek Zeno
Paradox to describe a peculiarity of the mound
formation~\cite{Elkinani-I1994,Elkinani-I1993}. In thin film growth, the
Zeno Paradox 
describes a situation where narrow trenches between mounds get so narrow,
that the probability for an atom or a molecule to land in the trench becomes increasingly
smaller as the trench width decreases. After the unlikely event
of a molecule entering the trench, the now narrower trench will have an
even smaller probability to be filled by further molecules, and thus stays
open. For this to be observable, a high step edge barrier has to prevent
molecules landing on higher lying terraces from descending onto lower terraces
deeper in the trench. As a result, the mounds get higher and higher but
would never coalesce thus the substrate would not be covered completely. 

A detailed analysis of the involved atomic or molecular diffusion processes
shows that in fact the time scales for the different basic processes
play an important role~\cite{Krug2000}. First, there is the traversal time 
\begin{equation}
   \tau_{tr}\approx A/\nu
   \label{equ:traversaltime}
\end{equation}
a particle needs to
visit all the sites on an island. Here, $A$ corresponds to the size of the
island (measured in lattice sites). The flux of incoming particles $F$ determines the time
\begin{equation}
   \Delta t=1/\left( FA \right)
   \label{equ:deposition-time}
\end{equation}
between the arrival of the particles forming the film. Finally, the residence
time~\cite{Krug2000a,Krug2000,Rottler1999,Heinrichs2000} 
\begin{equation}
   \tau=\frac{aL^2}{\nu}+\frac{bL}{\nu'}
   \label{equ:residencetime}
\end{equation}
describes the time a diffusing particle spends on an island with a
characteristic size $L$ ($L$ is the island's circumference). Here,
$a$ and $b$ are geometry dependent constants. While, the first term is on the order
of the residence time \eref{equ:residencetime}, the second term accounts for
the increase in residence time due to the step edge barrier. The ratio
$\alpha=\nu'/\nu$ of
the hopping rates for on terrace jumps~\eref{equ:hopping-island}
and
\begin{equation}
   \nu'=\nu_0 \e^{-E_S/(k_BT)}
   \label{equ:hopping-edge}
\end{equation}
for step edge crossings, can be used to
obtain the additional step edge barrier $\Delta E_{ES}$ between the barrier
for on island diffusion $E_D$ and the barrier for step crossing $E_S$. It is
crucial to realize at this point that $\nu'$ and therefore also
$\Delta E_{ES}$ can only be
effective values for interlayer mass transport. Different edge
terminations can in fact have very different hopping rates and
barriers. It is important to note that of all involved processes the one with
the smallest energy barrier will be the dominating
one, provided that the morphology connected with it is
occurring frequently~\cite{Teichert1994}. A very good review on mound
formation can be found in ref.~\cite{Michely2004}. 

\begin{figure*}[tb]
   \includegraphics[width=16cm]{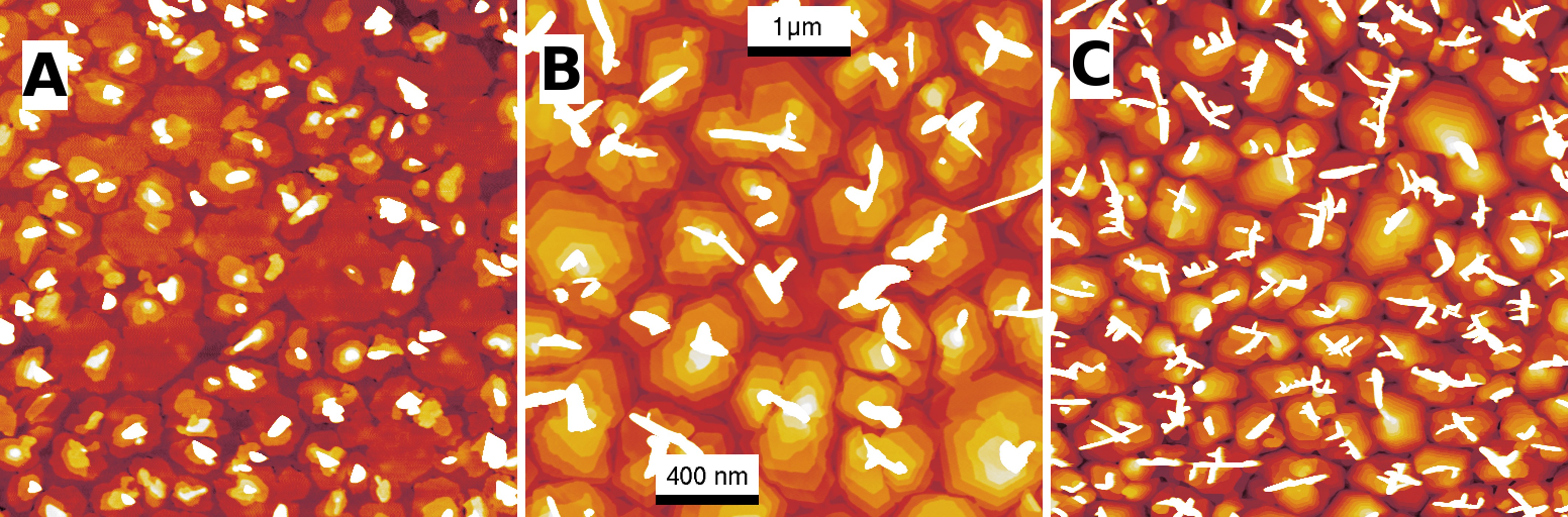}
   \caption{para--Sexiphenyl mound formation on sputtered mica.
      The evolution of the film morphology with growing film thickness
      (4\,nm, 10\,nm, and 30\,nm) can be seen. AFM image size:
      5\,\textmu{}m; z--scale: 20\,nm, 35\,nm, and 50\,nm.
      \reprintaaas[Adapted]{Hlawacek2008}} 
      \label{fig:mound-morphology}
\end{figure*}

\begin{figure}[tb]
   \begin{center}
      \includegraphics[width=8cm]{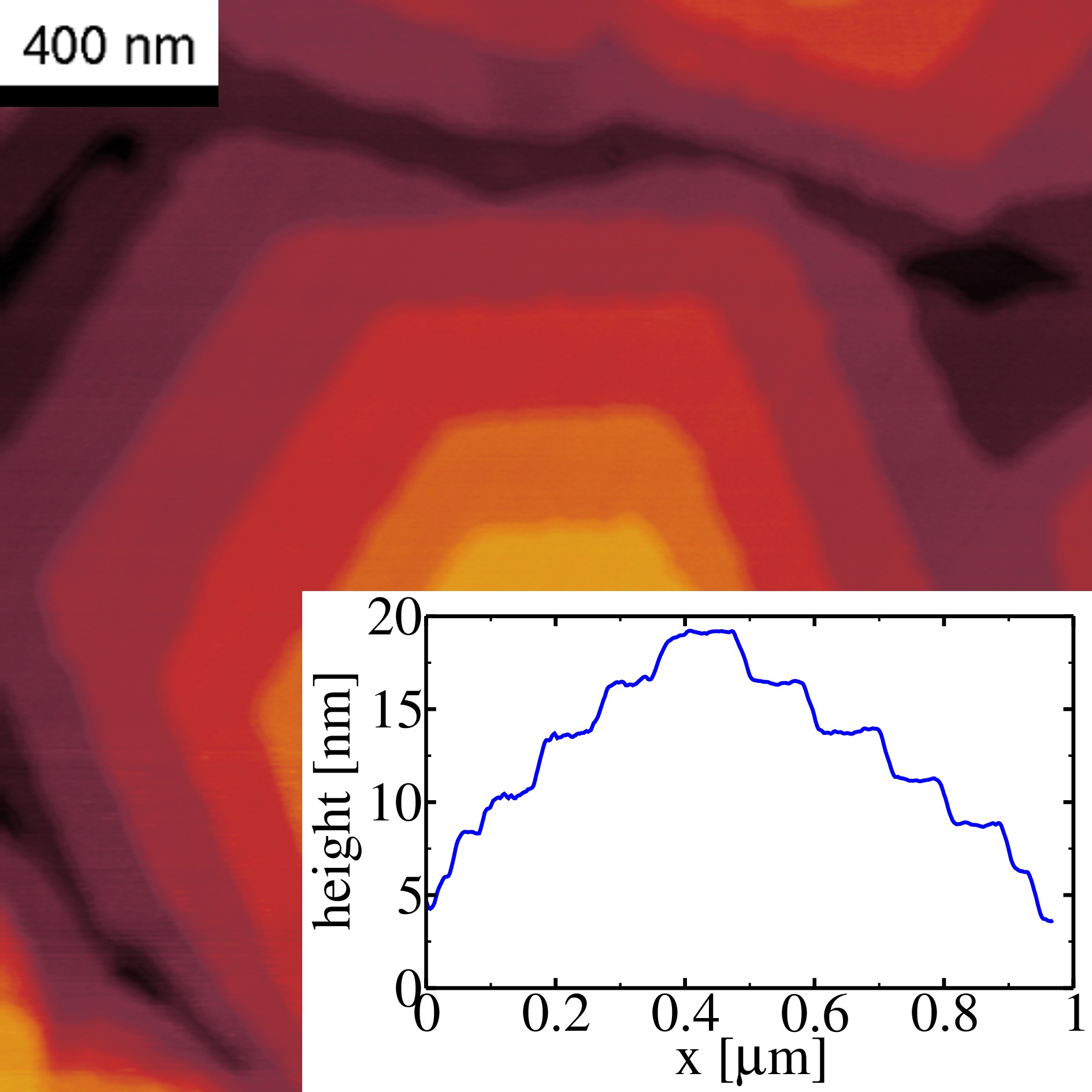}
   \end{center}
   \caption{Individual hexagonal shaped 6P mound and corresponding cross
   section. Single 2.7\,nm high terraces can be identified.}
   \label{fig:mound}
\end{figure}

Prototypical examples of mound formation
in thin films formed by rodlike molecules can be found in particular for
pentacene~\cite{Stadlober2006,Zorba2006} and
para--Sexiphenyl~\cite{Hlawacek2008}. Figure~\ref{fig:mound-morphology} gives
an overview on the morphology of growth mounds formed by 6P on a sputtered
mica surface at 300\,K.
Careful sputtering of the crystalline mica(0001) surface destroys the
symmetry of the surface and results in an disordered surface with an
unchanged chemical composition~\cite{Hlawacek2008,Frank2007}. This
modification
reorients the otherwise flat lying molecules into an upright orientation.
With increasing film thickness pronounced mounds start to form on the mica
surface. As can be seen from figure~\ref{fig:mound} and the corresponding
cross section the mounds are formed from upright standing molecules and
exhibit an irregular hexagonal shape. The steep trenches and the change in
curvature of the
mound slope that characterizes this morphology are an experimental
verification of the Zeno Paradox described by Elkinani and
Villain~\cite{Elkinani-I1993}. 

An analysis of this growth
behaviour~\cite{Krug2000,Politi1997,Kalff1999} allows to extract the
step--edge or Ehrlich Schwoebel barrier active in such a 
system~\cite{Hlawacek2008}. The step edge hopping rate in the system can be obtained
from the top terrace diameter~\cite{Krug2000}
\begin{equation}
   l\propto\left( \frac{\nu'}{F} \right)^\frac{1}{5}.
   \label{equ:topterrace}
\end{equation}
The size of the top terrace is limited by the fact that for terraces larger
than $l$ nucleation will occur on top of it making it the second to top
terrace. For the above presented film, $l$ has been measured to be
40\,nm$\pm$20\,nm~\cite{Hlawacek2008}. The probability for such a nucleation
event is related to the ratio between
$\Delta t$~\eref{equ:deposition-time} and the residence time
$\tau$~\eref{equ:residencetime}. The hopping rate for the on--terrace
diffusion can be obtained from kinetic nucleation theory~\cite{Venables1984}
which relates the nucleation density $N=1/\lambda^2$ to the hopping rate $\nu$ and the
flux $F$ via \eref{equ:scaling}.
%
%
The average island distance $\lambda$ is measured to be
1\,\textmu{}m~\cite{Hlawacek2008} for the
film presented in \fref{fig:mound-morphology}.
The relevant time scales here are again $\Delta t$~\eref{equ:deposition-time}
and the traversal time $\tau_{tr}$~\eref{equ:traversaltime}. The fact that the so
obtained value for $\Delta E_{ES}=0.67$\,eV is 30 times higher than the barrier
for 6P diffusion on top of a 6P(001) terrace (0.02\,eV determined by
molecular dynamics calculations~\cite{Hlawacek2008}) gives rise to the pronounced
mound formation in this system. Such high barriers are not uncommon for
organic systems. Fendrich et al. report 0.78\,eV for the Ehrlich--Schwoebel
barrier of flat lying 3,4,9,10-perylene--tetracaboxylic--dianhydride (PTCDA)
on PTCDA(102) using empirical potentials and the nudged elastic band
method~\cite{Fendrich2007b}. For both systems---6P and PTCDA---the
calculated diffusion barrier on the terraces is smaller by at least one
order of magnitude.

\begin{figure}[tb]
   \begin{center}
      \includegraphics[width=8cm]{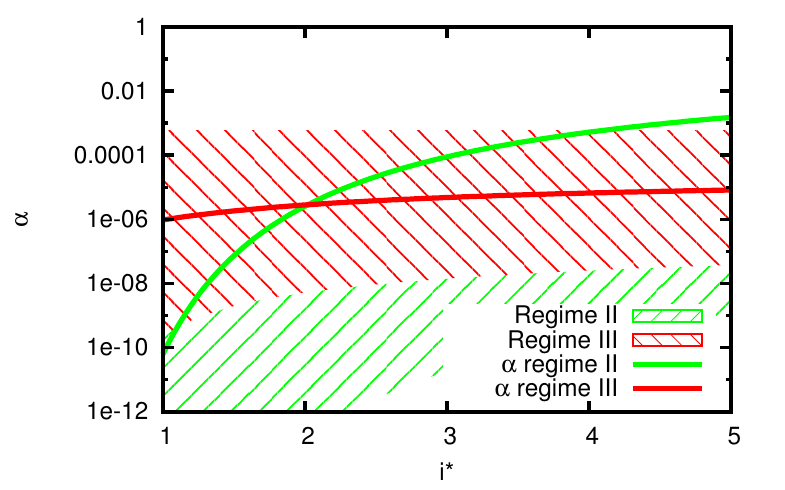}
   \end{center}
   \caption{Probability $\alpha$ for step edge crossing for different
      $i^*$. Solutions for regime II (green line) are only valid for
      $i^*=1$. The curve exceeds the valid range (green hatched area) for a
      larger critical nucleus. Solutions for regime III (red hatched area)
      are well within the bounds
      $\Gamma^{-\delta_1}\ll\alpha\ll\Gamma^{-\chi/2}$.}
   \label{fig:alpha-i-mounds}
\end{figure}

However, further careful analysis shows that the above used method
is only valid for the case of $i^*=1$. As we have seen in
\sref{sec:nucleus:obtainig} (\fref{fig:clusters}) this is rarely the
case in organic systems~\cite{Tumbek2012,Potocar2011}. Following the arguments
of~\cite{Krug2000a} we obtain 
\begin{equation}
   l\sim\Gamma^{\gamma'}\alpha^{\mu'}
   \label{equ:Krug2000a-mounds}
\end{equation}
with $\Gamma\equiv\nu/F$. Depending on the
detailed balance of the above presented time scales
\eref{equ:traversaltime}--\eref{equ:residencetime} one arrives at one of four possible regimes. For
regime I where $\alpha\ll\Gamma^{-1}$ we obtain pure Poisson growth. Regime
II extends from $\Gamma^{-1}\ll\alpha\ll\Gamma^{-\delta_1}$, where
$\delta_1=i^*(2i^*-1)/(2i^*(i^*+1)+2)<1$ for all $i^*$. The exponents
in~\eref{equ:Krug2000a-mounds} are than given by
\begin{equation}
   \gamma'=\mu'=\frac{i^*}{3i^*+2}.
   \label{equ:regimeII-mounds}
\end{equation}
For larger values of $\alpha$ but still smaller than
$\Gamma^{-\chi/2}$ the exponents take the form
\begin{equation}
   \gamma'=\frac{i^*}{i^*+3},\qquad\mu'=\frac{i^*+1}{i^*+3}.
   \label{equ:regimeIII-mounds}
\end{equation}
For $\alpha>\Gamma^{-\chi/2}$, we enter the regime of weak barriers and the
mound cross section starts to deviate from the above presented wedding cake
shape. \Fref{fig:alpha-i-mounds} plots the evolution of $\alpha$ in regime II
and III for
different $i^*$ together with the extent of their validity.
As one can immediately recognize, regime II is only valid for the case of
$i^*=1$. The value of $\alpha\approx2\times10^{-10}$ can be expressed in
terms of the step edge barrier using~\eref{equ:hopping-island}
and~\eref{equ:hopping-edge}. The exponential prefactors $\nu_0$ and
$\nu'_0$ are assumed to be equal. As the current regime II reproduces the
scenario used in~\cite{Hlawacek2008} we obtain a
similar result of $\Delta E_{ES}=0.58$\,eV. The difference is attributed to the
factor of probability which has a small effect on the final result and is
neglected for this overview. This result is only valid if the pair dissociation
time $\tau_{dis}\gg\tau_{tr}^2/\tau$, which is not the case for $i^*>1$. We
do find valid solutions
for $\tau_{dis}\ll\tau_{tr}^2/\tau$ (this regime III in~\cite{Krug2000a}
and~\cite{Tersoff1994} corresponds to regime I
in~\cite{Krug2000}). A smaller barrier of $\Delta E_{ES}=0.36$\,eV is obtained in
this regime. From \fref{fig:alpha-i-mounds} one can
see that there is a weak dependency of $\alpha$ on $i^*$. However, the
change is within the error bar of the experimental data used as input for
the calculation. As discussed above, for values of $\alpha$ above or below
the marked regions II and III one enters pure Poisson growth or the weak
barrier regime. Please note that
for larger $i^*$, regime III splits into a fluctuation and a mean field regime
with identical scaling exponents, and the evolution presented in \fref{fig:alpha-i-mounds} has to be
carefully reviewed for $i^*>2$. The difference being that for a small
number of involved particles the common mean field approach is not valid.
The above presented method---taking into account the statistical nature of the
initial nucleation---is needed in the case of small numbers. The complete
phase diagram of $-\ln\alpha/\ln\Gamma$ vs $i^*$ can be found
in~\cite{Krug2000a}. However, taking into account the size and
dimensionality of the building blocks we expect the fluctuation governed
regime to be valid for larger $i^*$ values than for atomistic processes.

\begin{figure}[tb]
   \begin{center}
      \includegraphics[width=8cm]{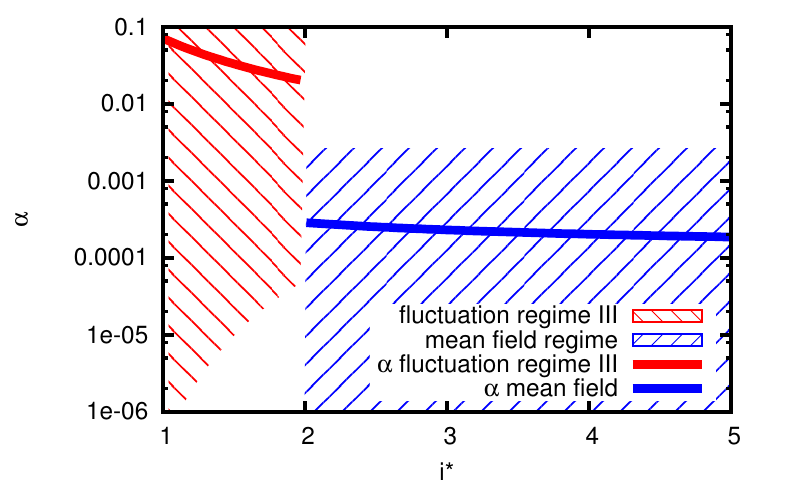}
   \end{center}
   \caption{Probability for step edge crossing for different $i^*$. The
   solution in regime III is plotted in red together with the range of
   applicability. For $i^*>2$ the fluctuation controlled method becomes invalid
   and the mean field approach (blue) becomes valid.}
   \label{fig:alpha-i-second}
\end{figure}

However, the barrier---even for the same edge---must not necessarily be constant during thin film
growth. Two things have been revealed for the initial growth of 6P on
sputtered mica (see \fref{fig:6PonMica}(d)). First, for the given film
thickness too few second layer islands have
nucleated. Second the island are only 2\,nm high, indicating a larger tilt
angle for the molecules. As a result a value of $\Delta E_{ES}=0.26$\,eV is reported for the
first layer ESB in~\cite{Hlawacek2008} using the method presented
in~\cite{Krug2000}. However following the arguments presented above,
this calculation and the size of the resulting step edge barrier has to be
carefully revisited. The analysis based on~\cite{Krug2000a} shows that only
for regime III a valid solution can be found. In regime III the critical
islands size for second layer nucleation
\begin{equation}
   L_c\sim\Gamma^\gamma\alpha^\mu
   \label{equ:critical-island}
\end{equation}
can be calculated using the exponents
\begin{equation}
   \gamma=\frac{\chi+i^*}{i^*+5}\text{, and} \qquad\mu=\frac{i^*+1}{i^*+5}.
   \label{equ:regimeIII-secondlayer}
\end{equation}
An additional problem in this calculation arises from the fact that in 
\begin{equation}
   f=1-\exp^{\frac{L}{L_c}^{k+2}}
   \label{equ:second_layer_number_density}
\end{equation}
which relates the second layer island fraction $f$ to the critical island size
for nucleation we find the exponent $k$.
This exponent $k$ is known to be 5 for fluctuation controlled nucleation
in regime II in case that  $i^*=1$~\cite{Krug2000}. By comparing (15) with (18) in~\cite{Krug2000a} and
(21) and (22) in~\cite{Krug2000} we obtain the general expression for regime
III $k=i^*+3$. The result of these considerations used in
\eref{equ:critical-island}, \eref{equ:regimeIII-secondlayer}, and
\eref{equ:second_layer_number_density} is plotted in
\fref{fig:alpha-i-second}. 
The red curve shows the expected $\alpha$ for
different $i^*$ in regime III. Although this curve ranges within its limits
$\alpha_X=\alpha^{1/(2i^*-1)}\gg\alpha\gg\alpha^{-1/3}\lambda^{-2/(3\chi)}=\alpha_F$
for small and large $i^*$,
the general condition for the validity of the fluctuation controlled regime
\begin{equation}
   \chi<\frac{2}{i^*+1}
   \label{equ:fl-valid}
\end{equation}
limits its extend to values of $i^*\le2$. For larger values of $i^*$ the
mean field approach from~\cite{Tersoff1994} becomes valid. Based on 
the expressions (8b) and (7) in~\cite{Tersoff1994} we obtain
\begin{equation}
   R_{c2}\sim\left[
   (i^*+5)\frac{L^2}{2\pi^3}4^{i^*}\left(\frac{1}{2}\right)^{i^*+1}\lambda^{2(i^*+2)}\alpha^{i^*+1}
   \right]^\frac{1}{i^*+5}
   \label{equ:TDT-Rc2}
\end{equation}
and 
\begin{equation}
   f=1-\exp^{-\frac{L}{L_c}^{m}}
   \label{equ:TDT-island-number-density}
\end{equation}
for the critical island radius~\eref{equ:TDT-Rc2} and the second layer
island fraction~\eref{equ:TDT-island-number-density}. For
the limit of high barriers---so that $\alpha\ll2/R_c$---the exponent has the
form $m=2i^*+6$. The result is plotted in blue in \fref{fig:alpha-i-second}. 
Based on this new analysis of the second layer nucleation we obtain a step
edge barrier $\Delta E_{ES}\approx0.1$eV in the fluctuation controlled regime for
$i^*\le2$. For larger $i^*$ values, as we deal with here, we obtain $\Delta E_{ES}\approx0.2$\,eV using
the mean field approach. The latter result has only a very weak dependence
on $i^*$. The value of $\Delta E_{ES}$ for second layer nucleation of at least 0.1\,eV is
substantially different from the above presented result obtained from the
Zeno Paradox and the overall mound shape of thicker films
($\Delta E_{ES}=0.36$\,eV). The observed change in molecular tilt angle for
the first few layers~\cite{Hlawacek2008} can explain this discrepancy. A
molecule crossing the step edge with its long molecular axis roughly
perpendicular, will bend over the edge while descending. As the molecular tilt
in the (001) plane gets smaller the necessary bending and thus the required
energy gets smaller, too.

The situation for a nucleus formed by flat lying molecules is more
complicated, since we can not estimate the order of the dimer dissociation time
from the finite dimer energy. In addition sterical hindering during the
nucleation starts to play a role and one enters the regime of
ALA~\cite{Tumbek2012}. In principle, the above
considerations can be extended into this regime. For such a situation the
exponent in \eref{equ:scaling} obtains the form
$\chi=2i^*/(i^*+3)$~\cite{Kandel1997}. In addition, the growth
laws---forming the foundation for the calculation---will play an important
role. 

When the molecule crosses the step edge it performs a
complicated sequence of twisting, rotating, and bending~\cite{Goose2010}.
All these processes are costly in terms of energy, and add to the final
barrier height the molecule has to overcome. However, the molecule will take
the pathway for which all contributions are added in such a way that the
final barrier will be minimal. In particular, for a molecule crossing the
step edge with the long axis roughly perpendicular to the edge, the bending
energy is a significant contribution. A decreased tilt of the molecular
backbone during the crossing lowers this energy term and consequently the
overall barrier height. 

The existing studies demonstrate that the prediction of barrier heights is
complicated and full of pitfalls. For the presented case, a good agreement
has been achieved initially between theory and experiment. The analysis
assumed a molecule that crosses the step edge with the long axis
perpendicular to the edge~\cite{Hlawacek2008}. However, after relaxing some
of the constraints used, other trajectories---with lower barriers (($\Delta
E_{ES}=0.34$\,eV))---involving difficult \emph{Fosbury Flop}--like movements
at the step edge were found~\cite{Goose2010}. The above presented new
analysis of the experimental data is in good agreement with the evolved
molecular dynamics simulations that predict a complicated step edge crossing
process and comparable barriers. Focused research in this direction is
important, as mound formation and layer dependent ESB values are common in
organic thin film growth. A level dependent ESB (e.g. as described above and
in~\cite{Hlawacek2008,Teichert2013}) often goes hand in hand with a change
in tilt angle of the molecular backbone. However, the proposed \emph{Fosbury
Flop}--like step edge crossing can not explain the experimentally observed
step edge barrier reduction. Provided the ESB becomes small enough the
initial layers can completely close~\cite{Teichert2013}. Other examples
include the growth of DIP on native SiO$_x$, which is characterized by a
transition from LbL growth to mound growth. This is explained by changes in
the interlayer mass transport~\cite{Zhang2009}. 

Care has to be taken when comparing experimental results with simulations
and DFT based calculations. It is important to realize that the
experimentally obtained barriers based on averaged mound shapes represent
the effective barrier in the entire film. Using well defined step edges (the
(100) in~\cite{Hlawacek2008,Goose2010}) allows detailed insight into the
dynamics of the step edge crossing for the specific facet. However, the
elongated hexagonal shape of the 6P islands (see \fref{fig:mound}) has at
maximum 2 of these (100) step edges. In addition, these two (the (100) and
the ($\overline{1}00$)) will also have different tilt directions (inward and
outward tilt) with respect to the top surface, which in turn are both
different from the tilt angle of the other unit cell facets (the vertical
(010) facet) and other possible step edges. This and other peculiarities of
molecular step edges (see also Appendix B in~\cite{Goose2010}) show how
difficult it is to make precise predictions of experimentally obtained
values for step edge barriers. 

\subsection{Growth of three-dimensional islands and fibers}

Although smooth films are usually preferred, the crystalline and
one--dimensional nano--fibers presented in \fref{fig:6PonMica}(a,b) are one out of
many examples of a useful non--smooth
morphology. It is important to remember that these anisotropic structures
grow from flat lying molecules. The previous section dealt with upright
standing molecules where no or only a weak anisotropy in the substrate plane
can be expected. In particular, the fact that blue
lasing~\cite{Andreev2006b,Quochi2005} has been shown for these fibers and
that they can be used as waveguides~\cite{Balzer2005} opens several 
possibilities for applications.

Two cases have to be separated here. While often three-dimensional fibers
grow directly on the substrate (Vollmer--Weber growth), in particular the
fibers found on mica grow on a metastable wetting layer
(Stranski--Krastanov growth). In both cases the molecule--molecule
interaction dominates over the molecule--substrate interaction. The
difference between these two types of interactions is large enough to
facilitate the rearrangement and reorientation of entire crystallites as
entities. 

\begin{figure}[tb]
   \begin{center}
      \includegraphics[width=8cm]{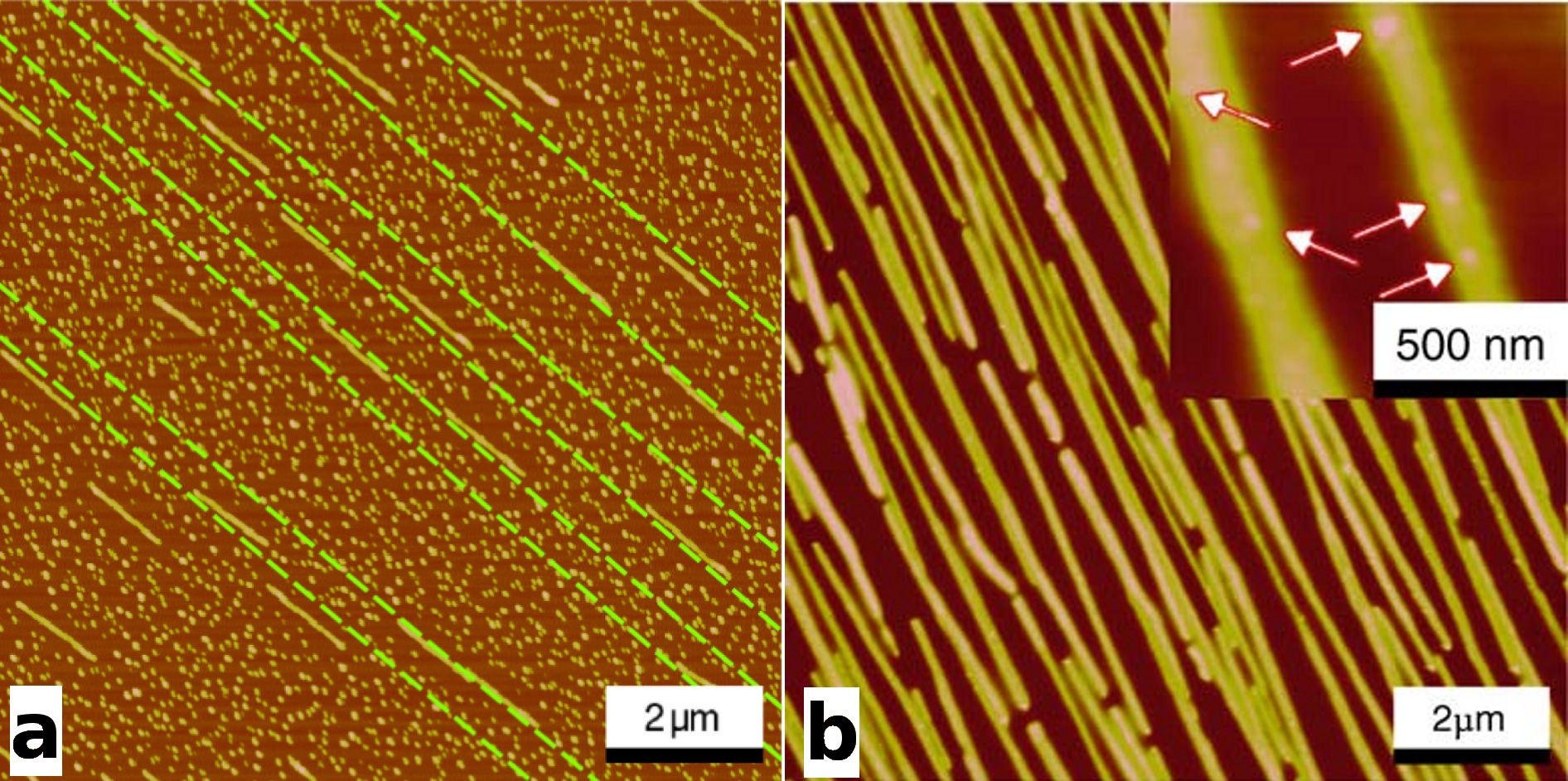}
   \end{center}
   \caption{AFM images of HWE grown 6P fibers on mica(0001). (a) Fibers---formed by a
   rearrangement process of several individual crystallites---line up on a
   dislocation network (indicated by green dashed lines) present in the wetting
   layer. (b) Detail of 6P crystallite chains with a length of several
   \textmu{}m, grown on clean mica under HV conditions. The high--resolution
   inset reveals the
   presence of small crystallites decorating the fibers.  Reprinted
   from~\cite{Teichert2006}. With kind permission from Springer Science and
   Business Media.}
   \label{fig:HWE-6P}
\end{figure}

The rearrangement of crystallites containing
more than 140000 molecules is observed during the HWE deposition of 6P onto
crystalline
mica(0001) at 360\,K~\cite{Teichert2006}. During the deposition of 6P, first
a wetting layer is formed. With increasing coverage crystallites grow on
this
wetting layer. However, after a critical amount of 6P has been deposited a
rearrangement takes place and fibers---formed by the already existing
crystallites---become the dominating morphological feature.
\Fref{fig:6PonMica}(a) shows such a chain of crystallites. The
particular arrangement of the individual chains with respect to each other
(see \fref{fig:HWE-6P}(a))
is explained by a strain relaxation mechanism at work in this
system~\cite{Teichert2006}.
The stress induced by the crystallites in the wetting layer leads to the
formation of a defect network (indicated by green lines in
\fref{fig:HWE-6P}(a)) that guides the rearrangement process of the
crystallites. It is important to realize at this point that during this
rearrangement process the crystallites move as entities. This relocation of
whole 6P crystallites on mica(0001) is possible due to the delicate balance
between the strong intermolecular interaction and the rather weak film
substrate interaction. Detailed x-ray diffraction (XRD) studies have revealed the epitaxial
relationship between 6P and the mica(0001)
surface~\cite{Simbrunner2011,Resel2003}. In particular, they have shown that
once the formation of needles sets in, the initially compressed spacing of
the $(11\overline{1})$ planes quickly relaxes towards the bulk
value~\cite{Andreev2004}. 

Recently, a bimodal size distribution for the crystallites on crystalline
mica has been observed. However, this behaviour for ultra--thin layers at a
slightly elevated temperature of 400\,K is only observed after exposing the
samples to ambient
conditions. Using TDS before and after exposing the sample to ambient
conditions as well as AFM revealed that the initial present wetting layer is
transformed into small crystallites. This second generation of smaller
crystallites
forms between the already existing fibers or chains of
crystallites.~\cite{Tumbek2012a}
For thicker films, the material from the wetting
layer is most likely captured by the large number of existing big
crystallites and fibers. In \fref{fig:HWE-6P}(b) a thick film
where long 6P needles have formed is presented. 
Several small second generation crystals are
visible in the inset of \fref{fig:HWE-6P}(b).

\begin{figure*}[tb]
   \begin{center}
      \includegraphics[width=16cm]{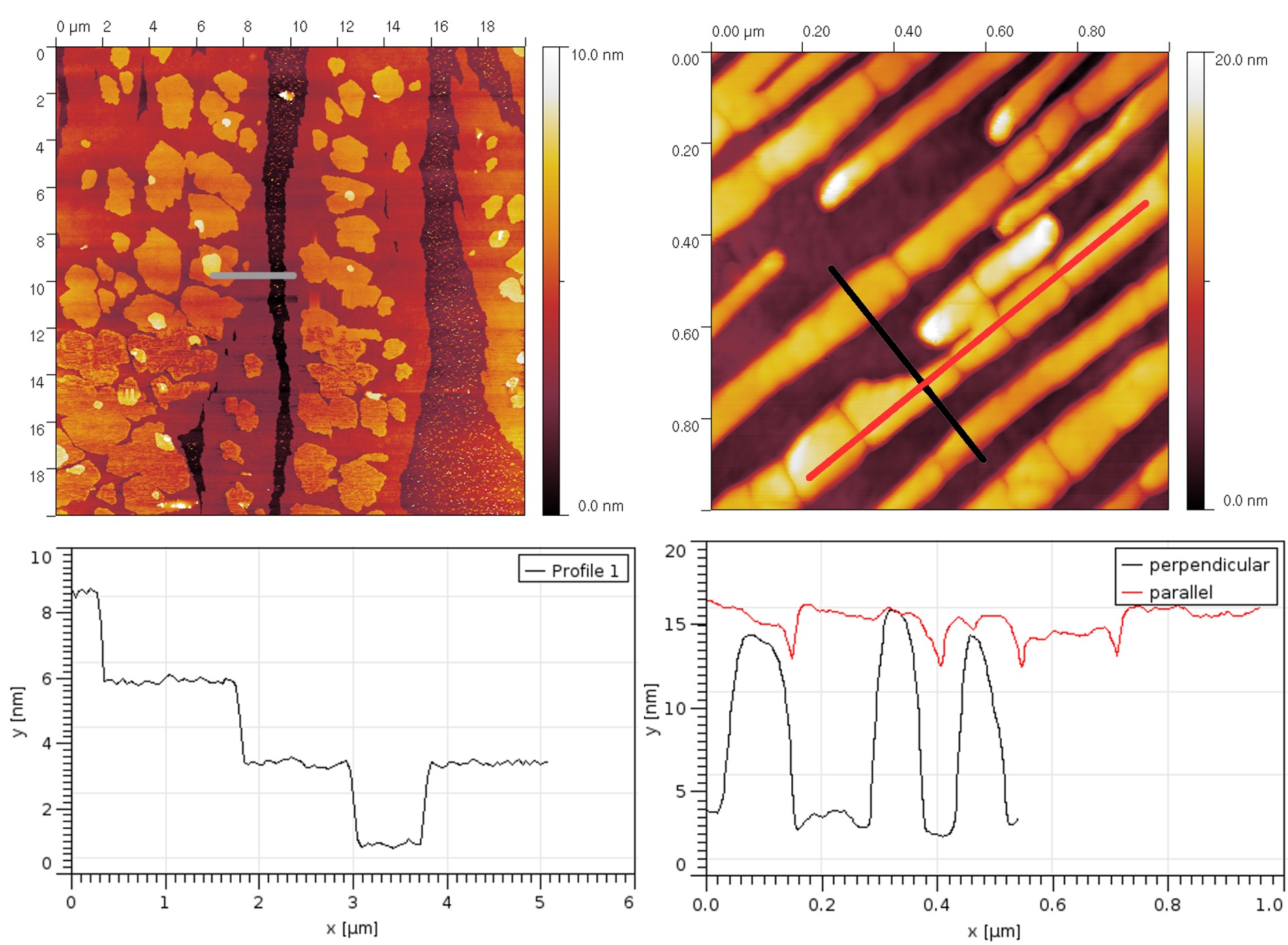}
   \end{center}
\caption{6P deposition on TiO$_2$\{110\}--(1$\times$1).\cite{Koller2004} (a) High
   temperature deposition at 400\,K leads to the formation of elongated
   islands formed
   by upright standing molecules. (b) Low temperature deposition at 300\,K
   results in the formation of long 6P fibers. Please note that the orientation
   of the structures has rotated by 90\textdegree{} from [001] to
   [1$\overline{1}$0]. (c,d) Cross--sections along the lines indicated in
   (a,b).}
   \label{fig:tio}
\end{figure*}

As we have seen for the case of crystalline mica versus sputtered mica, the
substrate plays an important role in determination of the molecular
orientation. However, also a particular surface reconstruction can provide 
an interesting growth template. The (1$\times$1) reconstruction of the
TiO$_2$\{110\} surface is
characterized by parallel
rows of protruding oxygen atoms. These rows run along the [001]
azimuth~\cite{Diebold2003}. The spacing of 6.5\,\AA{} is sufficient to
accommodate the width of a 6P molecule. Deposition of 1.3\,ML of 6P at
400\,K leads to the formation of large islands formed from upright standing
molecules presented in \fref{fig:tio}(a). The islands are separated by
trenches filled with small crystallites which are a few monolayers high. The
trenches run parallel to the oxygen rows along the [001] azimuth of the
TiO$_2$\{110\} surface~\cite{Hlawacek2005}. 
The islands themselves are
polycrystalline with four domains symmetrically spaced around the [001]
direction. With a size of only 300\,nm by 30\,nm, these domains are much
smaller than the several \textmu{}m large islands they form. The long axis
of these domains is also oriented along the [001]
direction~\cite{Resel2006a}. The reason for this growth behaviour is rooted
in the diffusion anisotropy present on this surface. The molecules can
easily diffuse along the [001] direction guided by the oxygen rows.
Analysing the width of the area in the trenches which is depleted from the
small crystallites, one arrives at a ratio for the anisotropy between the
diffusion along [001] and [1$\overline{1}$0] of 4 to
64~\cite{Berkebile2006}. This is clearly a property of the
TiO$_2$\{110\}--(1$\times$1) substrate surface since second and third layer islands show
an isotropic shape. All these structures formed by upright standing
molecules grow on top of a wetting layer of flat lying
molecules~\cite{Sun2010}. Such flat lying--to--upright transitions have been
observed for 5A on Cu\{110\}~\cite{Sohnchen2004} and other systems. For 
5A on Cu\{110\}, the transition involves a flat lying wetting
layer, which is followed by an intermediate layer having a herringbone structure
with the long molecular axis parallel to the substrate. For layers
thicker than 2\,nm an upright standing orientation is found in this system. 

\begin{figure}[tb]
   \begin{center}
      \includegraphics[width=8cm]{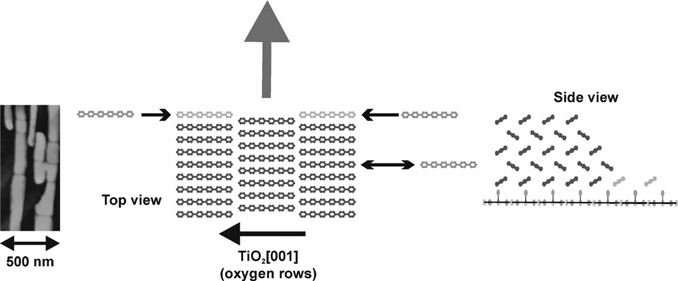}
   \end{center}
   \caption{Sticking anisotropy for 6P on TiO$_2$. The fibers shown in the
      AFM image (z--scale: 30\,nm) on the left grow along the
      [1$\overline{1}$0]. The sketch shows the molecular arrangement from
      the top and the side. The grey arrow indicates the growth direction.
      \reprintelsevier{Berkebile2006}{2006}} 
      \label{fig:tio-sticking}
\end{figure}

Lowering the growth temperature to 300\,K results in a complete change of
growth morphology, molecular orientation, and mesoscopic structure
orientation. The morphology presented in \fref{fig:tio}(b) is characterized
by long and high polycrystalline 6P fibers. It is important to realize that
these
fibers run parallel to the [1$\overline{1}$0] and thus perpendicular to the
oxygen rows and the trenches observed at higher temperatures. These fibers
are formed from flat lying molecules that have their long axis roughly
parallel to the substrate surface and are oriented
along the [001] direction of the TiO$_2$\{110\} surface. In addition to the
diffusion anisotropy active at high temperatures here the sticking
probability for molecules to be incorporated into existing fibers plays an
important role. The long side walls of the fibers are terminated by the
hydrogen atoms at the long end of the molecules. This has to be compared to
the short side of the fiber where the $\pi$ systems of the molecules are
exposed. It is clear that the sticking probability at the short end will be
substantially higher. Consequently the fibers will grow quickly along 
[1$\overline{1}$0] but slower in width. An illustration of the situation is
shown in \fref{fig:tio-sticking}.

An interesting mesoscopic approach to orient the fiber growth has been shown
by Madsen et al.~\cite{Madsen2009}. They used arrays of gold coated
micro--ridges. Tuning the ridge width and deposition temperature, 6P fibers
growing perpendicular to the ridges could be grown with a high yield.

\subsection{Layer--by--layer growth of lying molecules}

Although the above mentioned needle--like morphology might be useful for special
applications, a smooth interface is required for most applications. This is
in particular related to the fact that a lower number of defects at the
interface facilitates higher charge carrier
mobilities~\cite{Schumacher1982,Yan2010,Fritz2005,Steudel2004}. As we have
seen above films formed by upright standing molecules suffer in many cases
from high step edge barriers that will ultimately lead to mound
formation and rough interfaces. So far, efforts to
obtain Layer--by--Layer growth have led only to limited
success. 

Wu et al. have achieved five layers in the desired
LbL growth for the important case of 5A on SiO$_x$ by using 
SuMBD~\cite{Wu2009}. Using conventional OMBD,
Zhang et al. showed the strain relaxation driven transition from LbL to
rapid roughening after 5 layers for the plate--like molecule
DIP~\cite{Zhang2007}. In both studies, films formed from upright standing
molecules have formed. However, some success has been obtained for
flat lying molecules. 

Recently, Layer--by--Layer growth of flat lying molecules has
been obtained for 6P on the technological important substrate
graphene~\cite{Hlawacek2011,Lu2012a,Hlawacek2013}.
Graphene~\cite{Geim2007,Echtermeyer2008} can be used as a transparent,
flexible and highly conductive electrode for organic electronic
applications~\cite{Wang2009,Lauffer2008}. The combination of optical active
films of flat lying molecules on a transparent electrode materials is a
promising route to high efficiency OLEDs.

\begin{figure*}[tb]
   \begin{center}
      \includegraphics[width=16cm]{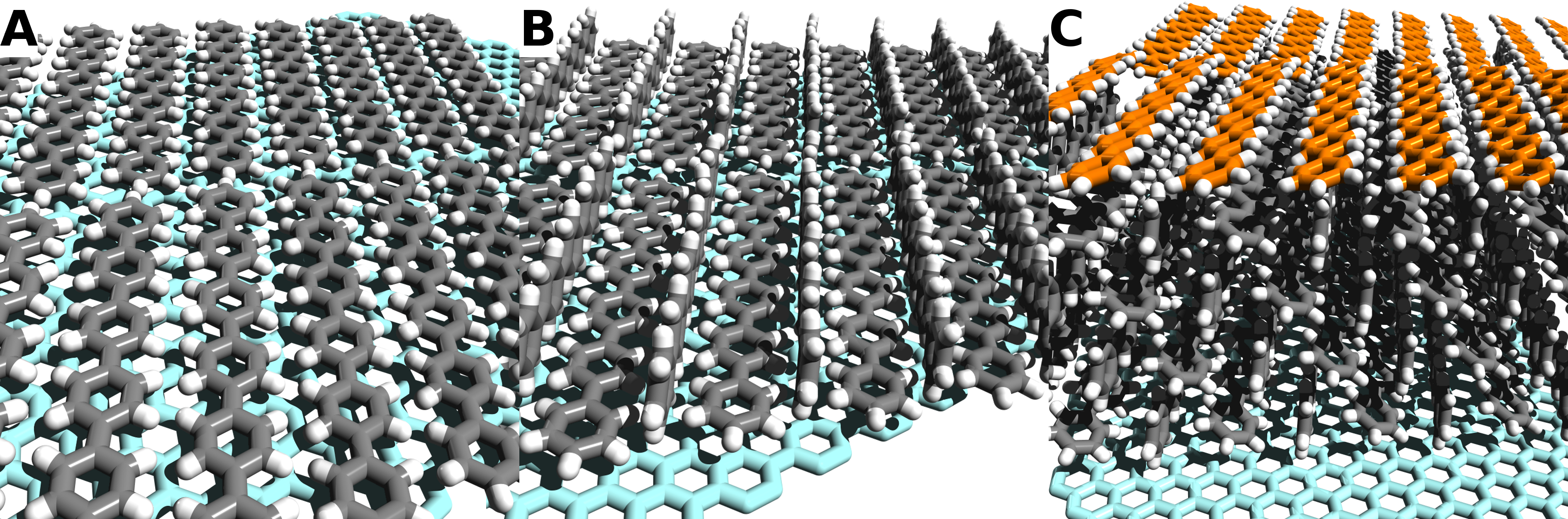}
   \end{center}
   \caption{Structural model of a 6P thin film grown on metal supported
      graphene (light blue hexagonal layer) at 240\,K. (a) The initial
      metastable layer is formed from flat lying molecules (grey carbon
      atoms) only. (b) After reaching a critical coverage, the structure
      changes to a bulk like molecular arrangement. This is achieved by
      tilting parts of the molecules on the long edge, as well as directly
      inserting molecules from the gas phase. (c) The film grows in LbL mode
      by a repetition of the previous two steps. The bulk of the thin film
      has a Baker--like structure~\cite{bakandfra93} and exposes the
      $(1\overline{11})$ plane to
      the underlying graphene substrate. The top most layer (orange carbon
      atoms) is not completed and shows a metastable structure consisting
      of flat lying molecules only. \reprintacs[Adapted]{Hlawacek2011}{2011}}
   \label{fig:LbL-model}
\end{figure*}

The formation of the 6P film in LbL mode at 240\,K has been
monitored using in--situ real time Low Energy Electron Microscopy (LEEM) and
micro Low Energy Electron Diffraction (\textmu{}LEED) for structural
characterization. The growth proceeds via a multi--step process that involves
the reorientation of a significant portion of already deposited
molecules~\cite{Hlawacek2011}. The process starts with the formation of a
metastable layer of exclusively flat lying molecules (depicted in
\fref{fig:LbL-model}(a)). 
With ongoing deposition, this highly
mobile initial layer~\cite{Hlawacek2011a,Hlawacek2013} transforms into a stable immobile
layer having a higher packing density and a bulk like arrangement of the
molecules (see \fref{fig:LbL-model}(b)). The structure of this stable layer
corresponds to the ($1\overline{11}$) plane of 6P. A similar growth process
for the first monolayer has been reported for 6P on
Au\{111\}~\cite{Mullegger2006}. However, on gold the growth at or above room
temperature results in three--dimensional growth. It is important to understand the
significance of the substrate for this process. Although earlier STM
studies of 6P on graphite~\cite{Wang2008} also report flat lying molecules,
the epitaxial relationship there is different to the one found on metal
supported graphene. Using empirical force fields and total energy
calculations it could be shown that indeed a different alignment of the long
molecular axis is favored on the two substrates~\cite{Hlawacek2011}. The
Layer--by--Layer growth process continues with the repetition of the above
two
steps~\cite{Hlawacek2011}. Every
additional layer starts with the formation of the metastable initial layer
of only flat lying molecules that transforms into a layer with the bulk
structure once a critical coverage is reached. \Fref{fig:LbL-model}(c)
shows the final film structure obtained by (\textmu{}LEED) for 4.5\,ML
coverage.  The achieved thin film structure bears the potential for high
efficiency OLED structures on a transparent and flexible
substrate~\cite{Hlawacek2011}. 

\section{Conclusion}

In the first part, we have discussed the nucleation and growth behaviour of rodlike molecules.
We presented several methods to determine the critical nucleus size. However, we
also showed that due to the non--zero--dimensional nature of the molecules, care has to
be taken when using formalisms originally introduced for atomic diffusion
processes. Many problems---originally identified in inorganic
systems---return in organic thin film growth. While phenomena like
attachment limited aggregation have been observed in a few inorganic systems,
they are encountered
on a regular basis in organic systems. However, the biggest difference is
due to the reorientation processes involved in the growth of films formed by
upright standing molecules. Initial insight is gained mostly by using
computational methods as the actual processes are difficult to monitor
experimentally. The most difficult question to answer is related to the
definition of the critical nucleus. Is an immobile cluster of flat lying
molecules that finally nucleates a
film formed by upright standing molecules the critical nucleus in a strict 
homoepitaxial sense? Although, this is to a certain level a semantic
question, one has to realize that most experimental techniques to
determine $i^*$ are insensitive to the orientation of the molecules.
Consequently, this problem needs to be discussed when interpreting the
results.

Furthermore, the molecular orientation plays a crucial role in defining the
efficiency of organic electronic devices. We have shown several ways to
influence the
orientation. We pointed out that substrate order and defects plays a crucial role for
switching from films formed by flat lying molecules to films made from
upright standing molecules.

In the second part of this review we focused on diffusion processes that define the final
film morphology. We extensively discussed mound formation in the presence of
an effective Ehrlich--Schwoebel barrier and in particular
the theoretical problems that arise when the critical nucleus becomes large.
However, as expected a
larger $i^*$ will not lead to a decrease of the involved step edge barriers.
Two additional remarks have to be added to this discussion. First, given the
size and the additional rotational degrees of freedom of molecules compared
to atoms, the transition from the fluctuation determined regime to the mean
field regime probably occurs at larger $i^*$ than in inorganic growth.
Second, we briefly discussed how to extract the critical nucleus sizes
for attachment limited aggregation (ALA) and diffusion limited aggregation
(DLA). The latter is also observed in organic growth and leads to the
formation of ramified islands. In the case of ALA, less molecules arrive at
the lower step edge thus reducing the upward diffusion flux. For reasonable
barrier heights this should facilitate LbL--like growth. In the case of DLA,
the number of kink sites increases substantially, creating more low--barrier
pathways over the step edge. While for ALA edge diffusion is in principle
possible, this mechanism is not effective in DLA. Consequently, more and
longer undesired domain boundaries are expected for DLA growth. The process
of mound formation and its relation to Layer--by--Layer growth gets further
complicated by the fact that the ESB can be layer dependent. In organic
epitaxy often a change of molecular tilt angle is the root cause for this change
of the barrier height and the growth mode.

Self--organization of the formed nano--structures can effectively be controlled
by balancing the anisotropies present within the growth system. This
includes---but is not limited to---diffusion and sticking anisotropy as well as
the anisotropy of the substrate on an atomic, but also mesoscopic length
scale. Also wetting
layers play an important role for many organic systems.  They are not
necessary stable under ambient conditions. However, rearrangement processes
mediated by the wetting layer---like the one observed for the crystallite
chain formation---can therefore only happen during (U)HV growth. No
change of morphology by such a process is possible once the wetting layer
has dissolved. 

\begin{figure}[tb] 
   \begin{center}
      \includegraphics[width=8cm]{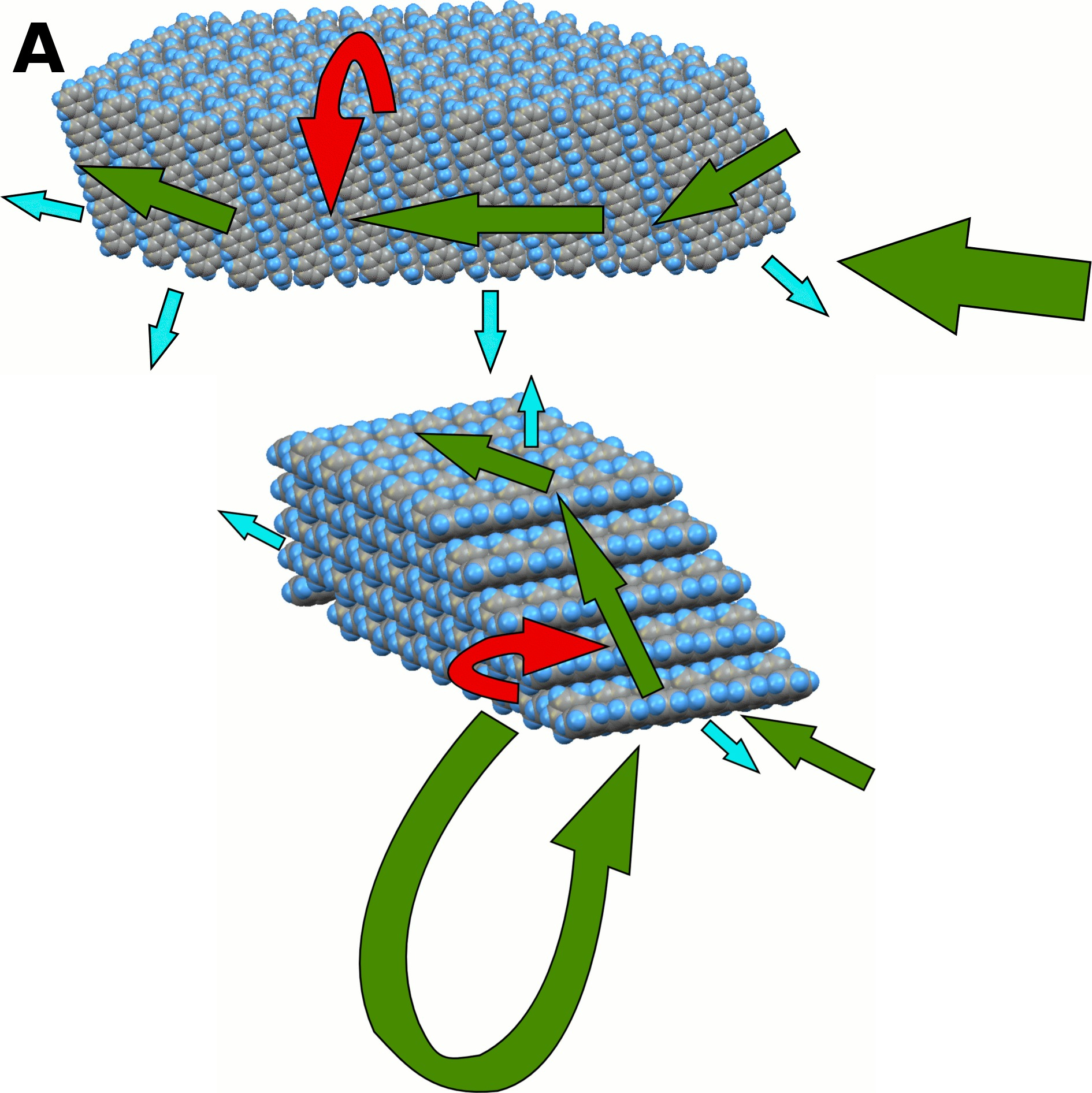} 
   \end{center} 
   \caption{Sketch highlighting probable (green) and less probable (red)
      diffusion paths for rodlike organic molecules. The same diffusion
      paths with nearly identical barriers (ignoring the substrate effect)
      exist for structures formed from either upright or flat lying
      molecules. While for upright standing molecules the red process is
      forbidden due to a high ESB, this process can be circumvented for
      structures formed by flat lying molecules.} 
   \label{fig:pathes} 
\end{figure}

In general, the anisotropy of the molecules leads to an anisotropy between
the different diffusion processes. However, depending on the actual
orientation of the molecules with respect to the substrate the diffusion
pathways change their meaning. The path indicated by the red arrow in
\fref{fig:pathes}(a) for the step edge crossing has a low probability and
results in the undesired mound formation.  
However, for the case of flat lying molecules (\fref{fig:pathes}(b)), the same
diffusion process can actively be avoided. The small sticking
probability at---what is now---the side of the fiber, allows the molecules
to circumvent the red diffusion process by returning into the gas phase and
reattachment at the small end. The consequence is the often observed fiber growth perpendicular
to the long molecular axis. The same analogy holds for edge
diffusion  (upright standing molecules) and up hill diffusion (flat lying
molecules). 

Finally, we presented results of LbL growth of rodlike molecules and
illustrated the often complicated rearrangement process occurring during
growth of organic thin films (e.g. as for the above described
low--temperature growth of 6P on
graphene). Similar as above a deviation from the bulk
structure goes hand in hand with a change of growth mode---in this case from
three-dimensional needles to Layer-by-Layer growth of flat lying molecules. The involved
metastable islands exhibit an interesting diffusion behaviour which is
mediated by a delicate interplay of strains in film and substrate. 

Considering the above presented information, two issues become immediately
evident. Forcing the molecules to deviate from their desired bulk structure
results in new and interesting growth phenomena and presents a viable
route for controlling the film morphology. Secondly, the level of
understanding of organic thin film epitaxy has increased dramatically within
the last decade. However, there is still a
large number of open questions. Having
identified these questions, dedicated experiments---supported by computational
methods---have to be designed to answer the existing challenges.

\ack

We grateful acknowledge all the people who were involved in the original
research funded by the Austrian Science Fund (FWF) under Projects No.
P19197, S9707, and S9714. In particular Claudia Ambrosch-Draxl, Andree
Andreev, Stephen Berkebile, Paul Frank, Stefan Lorbek, Stefan M\"ullegger, Bene Poelsema, Peter Puschnig, Mike Ramsey,
Roland Resel, Serdar Sariciftci, Quan Shen, Helmut Sitter, Raoul van Gastel,
and Adi Winkler. In addition we want to express our thanks to Andrei
Kadashchuk and Horst-G\"unter Rubahn. Finally we want to thank Harold J. W.
Zandvliet for making this manuscript possible.

\bibliography{Hlawacek-Diffusion-rod-like-molecules}
\bibliographystyle{iopart-num}


\end{document}